\documentclass[prb,superscriptaddress,floatfix,nofootinbib]{revtex4-2}
\usepackage[colorlinks=true,linkcolor=blue,citecolor=blue]{hyperref}
\usepackage{amsmath,amssymb}
\usepackage[dvipdfmx]{graphicx}
\usepackage{float}
\usepackage{comment}
\usepackage{xcolor}
\usepackage{bm,latexsym,amsmath,amssymb,amsfonts,mathrsfs}
\usepackage{bbold}
\usepackage{ascmac}
\usepackage{physics}
\usepackage{color,float}
\usepackage{enumerate}
\begin{document}
\title{
Anomalous Bulk Current in Quantum Hall Systems with an Expanding Edge
}

\author{Yuuki Sugiyama}
 \email{sugiyama.yuki@issp.u-tokyo.ac.jp}
 \affiliation{Institute for Solid State Physics, The University of Tokyo, Chiba 277-8581, Japan}
 
\author{Tokiro Numasawa}
 \email{numasawa@issp.u-tokyo.ac.jp}
 \affiliation{Institute for Solid State Physics, The University of Tokyo, Chiba 277-8581, Japan}
 
\begin{abstract}
Understanding topological phases of matter is essential for advancing both the fundamental theory and practical applications of condensed matter physics.
Recently, a theoretical framework for a quantum Hall system with an expanding edge state was proposed [Phys. Rev. D {\bf 105}, 105009 (2022)], revealing the existence of an energy flux analogous to Hawking radiation on the edge.
Motivated by this work, we extend the analysis to a model of a $(2+1)$-dimensional spacetime that includes both the bulk and edge regions.
Due to the presence of bulk and edges, we demonstrate that the covariant form of the gravitational anomaly appears on the edge via the anomaly-inflow mechanism.
Then, we investigate the energy flux on the edge from the viewpoint of gravitational anomalies, such as covariant gravitational and Weyl anomalies.
We also find that, due to the conservation of energy and momentum in the entire system, the presence of anomalous currents on the expanding edge induces non-trivial currents in the bulk originating from the expansion of the edge.
%{\color{red}Our results may...}
\end{abstract}

\maketitle

\section{Introduction}
The study of topological phases of matter has become a central theme in condensed matter physics.
Topological matters are expected to be the basis for state-of-the-art technologies such as quantum computers~\cite{Nayak2008, Bernevig2013, Moore2021}.
Unlike conventional phases distinguished by local order parameters, topological phases are characterized by global, quantized invariants~\cite{Wen1995, AZ1997}. 
%that remain robust under local perturbations~
Among these, the quantum Hall effect stands as a prototypical example: the quantized Hall conductance, expressed in units of fundamental constants, is intimately related to a topological invariant known as the Chern number~\cite{TKNN1982, Kohmoto1985, Hatsugai1993}. 
This discovery unveiled a deep connection between topology and physical observables and has since inspired the exploration of various novel phases, such as topological insulators and topological superconductors~\cite{Schnyder2008, Kitaev2009, Ryu2010, Bernevig2013}.
Quantum Hall systems thus provide an ideal setting to investigate the rich interplay among topology, symmetry, and many-body quantum physics in a highly controllable environment.

The quantum Hall system has been thoroughly investigated in static, equilibrium settings, where its essential features are now well understood~\cite{Yoshioka2002}.
Most experimental studies to date have been conducted under static conditions: the electrons are confined within the bulk by an electrostatic potential generated by the surface of the host semiconductor, and the edge, defined by the boundary of this potential, remains stationary in time.
%The bulk is typically realized as a two-dimensional electron gas (2DEG) subjected to a strong perpendicular magnetic field, giving rise to Landau levels and resulting in incompressible quantum fluids at certain filling factors. 
The bulk system is a two-dimensional electron gas (2DEG) subjected to a strong perpendicular magnetic field, giving rise to Landau levels and resulting in incompressible quantum fluids at certain filling factors. 
In contrast to the gap in the bulk, the edge hosts robust, gapless excitations protected by topology, which are effectively described by chiral quantum field theories in $(1+1)$-dimensional spacetime~\cite{Wen1995}. 
These edge modes reflect the non-trivial topological character of the bulk via the bulk-boundary correspondence~\cite{Hatsugai1993}.
%but also offer a versatile platform for studying low-dimensional transport~\cite{}, quantum anomalies~\cite{Callan1985}, and interaction effects~\cite{Hamanaka2024}.
The separation between the gapped bulk and the %dynamically constrained yet highly non-trivial {\color{red}dynamical (static ?)} 
gapless edge structure makes quantum Hall systems a powerful arena for probing the fundamental links between topology and quantum field theory.

While most experimental investigations have focused on static edge structures, recent theoretical proposals and ongoing experiments have begun to explore dynamical edge phenomena~\cite{Hotta2022, Nambu2023, Yoshimoto2025, Kamiyama20221, Kamiyama20222, Jeong2025}. 
In particular, the concept of expanding edges has been introduced, in which the boundary of a quantum Hall droplet evolves over time due to the gradual relaxation of the confining electric potential, coupled with continuous electron injection into the bulk. 
This expansion influences the propagation of edge excitations and introduces intrinsic time dependence to an otherwise stationary setup. 
Remarkably, this configuration enables analog simulations of quantum field theory in curved $(1+1)$-dimensional spacetime. 
In our previous work~\cite{Hotta2022}, we formulated a field theoretical description of such systems on a curved background and showed that the expanding edge can effectively simulate two-dimensional dilaton gravity models. 
Within this framework, we demonstrated that the expanding region of the edge gives rise to a future event horizon akin to that of de Sitter space, accompanied by a Gibbons-Hawking temperature. 
This suggests the intriguing possibility of observing classical analogs of Hawking radiation in condensed matter systems.

The previous studies mainly focused on the analysis of Hawking radiation in the expanding edge region~\cite{Nambu2023, Yoshimoto2025}.
In Ref.~\cite{Nambu2023}, the authors argued that the quantum correlations between two detectors in the outer region decohere by detecting the Hawking radiation.
In Ref.~\cite{Yoshimoto2025}, they proposed a method to derive Hawking radiation in the expanding edge region using gravitational anomaly.
Since gravitational anomalies imply a violation of the conservation of energy and momentum, they imposed the anomaly cancellation on the expanding edge, following the method originally developed for a black hole spacetime~\cite{Wilczek2005, Iso20061, Iso20062}.
%As a result, they derived an energy flux corresponding to Hawking radiation in the outer region.
This approach led to the derivation of an energy flux corresponding to Hawking radiation observed in the outer region.

However, these treatments neglect a crucial aspect of quantum Hall systems—the existence of the bulk.
%since the quantum Hall system is composed of a bulk and its edge as boundaries, the existence of the bulk cannot be ignored.
Indeed, the chiral edge modes have a gravitational anomaly, which is an obstruction to putting the chiral modes on a curved $(1+1)$-dimensional spacetime~\cite{Witten1985}.
This means that in quantum Hall systems on a curved spacetime, we cannot forget about the existence of the bulk, though the bulk has an energy gap and contains no dynamics.
%\textcolor{blue}{though the bulk has an energy gap and contains no dynamics.}
%Following the idea of the anomaly-inflow, where the (gauge) anomaly on the edge cancels out the conserved charge originating in the bulk~\cite{Callan1985}, the anomaly on the edge should disappear throughout the entire quantum Hall system.
%Rather
In fact, the quantum Hall systems realize the anomaly-inflow mechanism \cite{Callan1985}, where the bulk contribution precisely cancels the boundary anomaly, ensuring overall energy-momentum conservation.
%In other words, it is non-trivial whether the gravitationale anomaly vanishes if we focus only on the edge region.
In other words, the stress tensor is conserved only when we combine the bulk and boundary contributions.
%Furthermore, experiments and numerical simulations revealed that when edge states are excited in the quantum Hall system with the expanding edge, the modes of the bulk are also excited~\cite{Jeong2025}.
%This result suggests that there is a coupling between the bulk and edge degrees of freedom.
Furthermore, recent experiments and numerical simulations~\cite{Jeong2025} have shown that excitations in the edge region also lead to excitations in the bulk, providing direct evidence of bulk–edge interaction.

In the present paper, we consider a $(2+1)$-dimensional quantum Hall system with the expanding edge.
In particular, we consider the role of the gravitational Chern-Simons term that should be obtained by integrating the bulk gapped modes \cite{Witten:2019bou}.
%\textcolor{blue}{In particular, we consider the role of the gravitational Chern-Simons term that should be obtained by integrating the bulk gapped modes \cite{Witten:2019bou}.}
By considering the gravitational Chern-Simons action in the bulk, we show in the general case that the edge gravitational anomaly is in covariant form~\cite{Cappelli2002}.
We also see that the gravitational anomaly is equal to the bulk energy-momentum tensor near the edge.
Then, we derive the energy flux corresponding to Hawking radiation in two ways.
Firstly, we solve the covariant gravitational anomaly equation directly and derive the energy flux formula.
The second method is to construct a solution of the covariant gravitational anomaly equation using the Weyl anomaly formula of the energy-momentum tensor obtained from $(1+1)$-dimensional conformal field theory~\cite{Polyakov:1981rd,Polchinski:1998rq,Nakahara:2003nw,Maldacena2018}.
We show that the energy flux formula is proportional to the Schwarzian derivative, and the current corresponds to the Hawking radiation at late time.
%Finally, assuming spatial profile functions, which characterize the boundary between the flat and curved regions, we compute the bulk energy-momentum tensor numerically.
Finally, assuming the spatial profile of the expanding $(2+1)$-dimensional spacetime, which characterizes the localized expansion near the edge region, we compute the bulk energy-momentum tensor numerically.
Then, we discuss the effect of the expanding edge on the bulk energy-momentum tensor.

This paper is organized as follows.
In Sec.~II, we consider a $(2+1)$-dimensional quantum Hall system with a curved spacetime edge region.
Then, we derive the anomaly equation on the edge from the viewpoint of the anomaly-inflow.
In Sec.~III, we briefly review the setup of the quantum Hall system with a time-dependent expanding edge presented in the previous studies~\cite{Hotta2022, Nambu2023, Yoshimoto2025}.
Solving the gravitational anomaly equation on the spacetime, we derive the energy flux analogous to the Hawking radiation.
Furthermore, we also compute the energy flux formula using the general form of the energy-momentum tensor with Weyl anomaly.
In Sec.~IV, we focus on the bulk energy-momentum tensor and discuss the effects of expanding edges for the bulk region.
Section V is devoted to the conclusion.
In Appendix A, we summarize the formulas in general $(1+1)$-dimensional spacetime, such as the metric, totally anti-symmetric tensor, and the Christoffel symbols, both in the Minkowski and null coordinates.
In Appendix B, we present a detailed computation of the coordinate transformation in the quantum Hall system with an expanding edge.
In Appendix C, we present the proof of the equivalence of Eqs.~\eqref{ppcomponent} and~\eqref{weylppcomponentc}.
In Appendix D, we give the formulas for the bulk energy-momentum tensor.
Throughout the present paper, we use the natural units $c=\hbar=k_{\text{B}}=1$.
We also adopt the following notations: the greek indices $\mu, \nu, ...$ taking on the value $\mu, \nu, ...=0, 1, 2$, while the latin indices $i,j, ...$ doing $i,j, ...=0, 1$.
 
\section{Gravitational Anomaly and Anomaly-Inflow in QHS with curved edge}
In this section, we consider a $(2+1)$-dimensional spacetime where one spatial component varies with time. 
Specifically, we focus on a quantum Hall system (QHS) with a time-dependent expanding edge, while the 
%\textcolor{blue}{deep inside of the} 
deep inside the bulk remains unchanged.
This setting represents a situation where the edge states of the QHS evolve dynamically. 
The physical motivation for this setup is to analyze how the energy-momentum tensor behaves under such time-dependent conditions in terms of quantum anomalies.

To systematically describe this system, we introduce a spacetime metric that captures the edge expansion and derive the corresponding generating functional for both the edge and bulk contributions.
We define the coordinates as $x^{\mu}=(vt, x, y)$, where $v$ denotes a velocity of the system, which is identified as edge current in quantum Hall systems.
In what follows, we set $v=1$.
The coordinate $t$ parametrizes the conformal time coordinate.
%, related to a laboratory time $\tau$ via $t=\int d\tau a(\tau)$.
%The scale factor $a(\tau)$ characterizes the expansion of the edge.
The coordinates $x$ and $y$ are comoving spatial coordinates.
In this setup, we consider a scenario where the $x$-component expands with time, which leads to a non-trivial metric structure near the edge of the QHS.
The bulk of the QHS is introduced in a half-plane $y<0$, and the edge state is defined on the plane $y=0$.
The metric near the edge of the QHS is given by
\begin{align}
ds^2
=g_{\mu\nu}dx^{\mu}dx^{\nu}
=a^2(t,x,y)\eta_{ij}dx^{i}dx^{j}+dy^2,
\quad
a^2(t,x,y)
=\big(\omega^2(t,x)-1\big)f(y)+1,
\end{align}
where $g_{\mu\nu}$ is the metric tensor, which describes the geometry in $(2+1)$-dimensional spacetime.
The function $\omega(t, x)$ characterizes the geometry of the edge expansion and $\eta_{ij}$ is the Minkowski metric in $(1+1)$-dimensions.
The function $f(y)$ controls the spatial profile near the edge and satisfies the two conditions: $df(y)/dy|_{y=0}=0$ and $f(y)|_{y=0}=1$.
These conditions ensure a smooth connection between the edge and the bulk region.
In the present paper, we will consider two specific cases of $f(y)$ (the detailed discussions, see Sec.~IV).

Near the edge of the QHS, the metric is approximated as
\begin{align}
ds^2
=
h_{ij}dx^{i}dx^{j}+dy^2
=
\omega^2(t,x)\eta_{ij}dx^{i}dx^{j}+dy^2+\mathcal{O}(y^2),
\label{edge}
\end{align}
where $h_{ij}(t,x)=\omega^2(t,x)\eta_{ij}$ is the metric tensor in $(1+1)$-dimensions.
Note that we used the Taylor series of the function $f(y)$ around $y=0$ and neglected the order $\mathcal{O}(y^2)$. 
Thus, the edge mode effectively experiences a conformally flat spacetime.

The total generating functional $W_{\text{tot}}$ in this $(2+1)$-dimensional system consists of two contributions:
\begin{align}
W_{\text{tot}}
=W_{\text{edge}}+W_{\text{CS}}.
\end{align}
Here, $W_{\text{edge}}$ is the generating functional of the edge mode in $(1+1)$-dimensional system, producing massless chiral fermion~\cite{Callan1985} and the variation $\delta W_{\text{edge}}$ is defined by the energy-momentum tensor 
$T^{ij}_{\text{edge}}$
as follows:
\begin{align}
\delta W_{\text{edge}}
=
\frac{1}{2}
\int d^2x \sqrt{-h} T^{ij}_{\text{edge}}\delta h_{ij},
%T^{ij}_{\text{cons}}\delta h_{ij},
\end{align}
where 
$h:=\det(h_{ij})$
is the determinant of the metric $h_{ij}$.
%This contribution, $W_{\text{edge}}$, governs the dynamics of the edge excitations.

The generating functional in the bulk is effectively described by a gravitational Chern-Simons (CS) action $W_{\text{CS}}$\footnote{Of course there is a Chern-Simons term for the electromagnetic fields, but in this paper we do not consider the electromagnetic responses and we omit it here.}, which is parity-odd and captures the topological nature of the quantum Hall effect.
%In particular, a low-energy effective theory of the bulk in QHS is the Chern-Simons theory.
The Chern-Simons term appears by integrating out %fields coupled with 
bulk gapped fermion fields in the path integral formalism (e.g.,~\cite{Witten1985}).
The Chern-Simons action in curved spacetime is described by gravitational Chern-Simons theory, and it takes the form:
\begin{align}
W_{\text{CS}}
=
\beta
\int_{\mathcal{M}}
\text{Tr}
\qty[
\Gamma \wedge d\Gamma
+\frac{2}{3}
\Gamma \wedge \Gamma \wedge \Gamma 
],
\end{align}
where $\text{Tr}$ is the sum of the spacetime indices $\mu=0, 1, 2$, respectively, and $\Gamma$ is the Christoffel symbols in a 3-manifold $\mathcal{M}$ with the metric $g_{\mu\nu}$.
Here, the coefficient $\beta$ is the coefficient $\beta={c_{-}}/{96\pi}$ with chiral central charge $c_{-}$.
The non-zero chiral central charge means chirality because it counts the difference in degrees of freedom between the left and right-moving modes.
In particular, since no backscattering process occurs at the edge of QHS, we always deal only with the left-moving mode.
Note that we used the notation of the differential form: 
$\Gamma=\Gamma_{\mu}dx^{\mu}$ being a matrix valued
%\textcolor{blue}{matrix valued} 
1-form 
and its exterior derivative
$d\Gamma=\partial_{\mu}\Gamma_{\nu}dx^{\mu}\wedge dx^{\nu}
=(\partial_{\mu}\Gamma_{\nu}-\partial_{\nu}\Gamma_{\mu})
dx^{\mu}\wedge dx^{\nu}/2$ being a 2-form with wedge product $\wedge$.
For 3-form, 
$dx^{\alpha} \wedge dx^{\beta} \wedge dx^{\gamma}
=\epsilon^{\alpha \beta \gamma} dx^0 dx^1 dx^2
=\epsilon^{\alpha \beta \gamma} d^3x$
with $\epsilon^{012}=1$
is satisfied.

In bulk with boundary theory, anomaly-inflow is a flow of conserved charges from the bulk to the boundary~\cite{Callan1985, Harvey2005, Jensen2013}.
As we will see below, this flow of the conserved charges modifies the conservation law at the boundary to a covariant form.
Let us first consider the transformation of the total generating functional %\textcolor{blue}{partition function? [It is a common name in hep-th but I'm not confident the right name...]}
$W_{\text{tot}}$ under the diffeomorphisms.
Because the entire system should conserve the energy and momentum, $W_{\text{tot}}$ must satisfy 
\begin{align}
\delta_{\xi}W_{\text{tot}}=0,
\label{diff}
\end{align}
where $\delta_{\xi}$ denotes the diffeomorphism variation originated from the infinitesimal transformation of the coordinates $x^{\mu} \to x^{\mu}-\xi^{\mu}(x)$ with 
$\xi^{\mu}(x)$ being the infinitesimal transformation parameter.
The diffeomorphism invariance guarantees energy and momentum conservation in the entire system.
%\begin{align}
%\nabla_{\mu}T^{\mu\nu}_{\text{tot}}=0.
%\end{align}
Eq.~\eqref{diff} leads to the following equation~\cite{Jensen2013}
\begin{align}
0=
\delta_{\xi}W_{\text{edge}}
+\delta_{\xi}W_{\text{CS}}.
\end{align}
The first term reads
\begin{align}
\delta_{\xi}W_{\text{edge}}
=
\frac{1}{2}
\int d^2x \sqrt{-h}T^{ij}_{\text{edge}}
%T^{ij}_{\text{cons}}
\delta_{\xi} h_{ij}
=
-\int d^2x \sqrt{-h}
\nabla_{i}T^{ij}_{\text{edge}} \xi_{j},
%T^{ij}_{\text{cons}} \xi_{j}
\end{align}
where $\delta_{\xi}h_{ij}=\nabla_{i}\xi_{j}+\nabla_{j}\xi_{i}$ in the covariant derivative $\nabla_{i}$ with $h_{ij}$.
Note that we neglected the boundary term originating from the partial integration because %we are interested in more than $(1+1)$-dimensions.
we can assume that $\xi$ only has support on a small region and vanishes at the boundary in $x$-direction.
%\textcolor{blue}{we can assume that $\xi$ only has a support on a small region and vanish at the boundary in $x$ direction}.
The second term $\delta_{\xi}W_{\text{CS}}$ satisfies the following equation~\cite{Bertlmann1997}:
\begin{align}
\delta_{\xi}W_{\text{CS}}
%=
%\beta\int_{\partial \mathcal{M}}\text{Tr}[vd\Gamma]
%=
%\beta\int d^2x \epsilon^{lm}\partial_{i}\xi^{k}\partial_{l}\Gamma^{i}_{mk}
=
-\beta\int d^2x \sqrt{-h}
h^{jk}\bar{\epsilon}^{lm}\partial_{i}\partial_{l}\Gamma^{i}_{mk}
\xi_{j}
\end{align}
with the covariant totally anti-symmetric tensor
$\bar{\epsilon}^{lm}
:=\epsilon^{lm}/\sqrt{-h}$
with $\epsilon^{01}=-\epsilon^{10}=1$
in $(1+1)$-dimensions.
Thus the tensor 
$T^{ij}_{\text{edge}}$ 
%$T^{ij}_{\text{cons}}$ 
satisfies the following anomaly equation:
\begin{align}
\nabla_{i}T^{ij}_{\text{edge}}
%\nabla_{i}T^{ij}_{\text{cons}}
=
-\beta h^{jk}\bar{\epsilon}^{lm}\partial_{i} \partial_{l} \Gamma^{i}_{mk}.
\label{consistent}
\end{align}
This form is not general covariant and is often called \textit{consistent gravitational anomaly} because it satisfies the Wess-Zumino consistency condition~\cite{Bertlmann1997,Hotta2009}.
This equation will be modified by considering the flow of conserved charges from the bulk.

A general variation of the Chern-Simons action with respect to $(2+1)$-dimensional metric $\delta g_{\mu\nu}$ is
\begin{align}
\delta W_{\text{tot}}
&=
\delta W_{\text{edge}}+\delta W_{\text{CS}}
\nonumber\\
\quad
&
=
\frac{1}{2}
\int d^2x\sqrt{-h}T^{ij}_{\text{edge}}\delta h_{ij}
%\int d^2x\sqrt{-h}T^{ij}_{\text{cons}}\delta h_{ij}
+
\frac{1}{2}\int d^3x\sqrt{-g}
T^{\mu\nu}_{\text{bulk}}
\delta g_{\mu\nu}
+
\frac{1}{2}
\int d^2x\sqrt{-h}P^{ij}\delta h_{ij}
\nonumber\\
\quad
&
=
\frac{1}{2}\int d^3x\sqrt{-g}
T^{\mu\nu}_{\text{bulk}}
\delta g_{\mu\nu}
+
\frac{1}{2}
\int d^2x\sqrt{-h}T^{ij}_{\text{cov}}\delta h_{ij},
\label{derivative}
\end{align}
where, in the second line, the energy-momentum tensor in the bulk $T^{\mu\nu}_{\text{bulk}}$ is
\begin{align}
T^{\mu\nu}_{\text{bulk}}
=
-2\beta
\Big(
\bar{\epsilon}^{\gamma\rho\mu}\nabla_{\gamma}R^{\nu}_{\rho}
+\bar{\epsilon}^{\gamma\rho\nu}\nabla_{\gamma}R^{\mu}_{\rho}
\Big).
\label{bulk}
\end{align}
Here,
$\bar{\epsilon}^{\gamma\rho\mu}
:=\epsilon^{\gamma\rho\mu}/\sqrt{-g}$
and
$R^{\nu}_{\rho}$
are the covariant totally anti-symmetric tensor and the Ricci tensor in $(2+1)$-dimensions.
The covariant derivative $\nabla_{\gamma}$ is defined on the spacetime described by $g_{\mu\nu}$.
The last term in the second line, we defined $P^{ij}$ and is explicitly given by
\begin{align}
P^{ij}
=-
\frac{\beta}{2}
(\delta^{i}_{p}\delta^{j}_{q}+\delta^{j}_{p}\delta^{i}_{q})
\Big(
\bar{\epsilon}^{rl}g^{mq}\nabla_{r}\Gamma^{p}_{ml}
+
\bar{\epsilon}^{pl}g^{mq}\nabla_{s}\Gamma^{s}_{ml}
-\bar{\epsilon}^{pl}g^{mr}\nabla_{r}\Gamma^{q}_{ml}
\Big)
=-\beta
\bar{\epsilon}^{pl}g^{rm}
T^{ijk}_{pqr}
\nabla_{k}\Gamma^{q}_{ml}.
\end{align}
In the second equality, we introduced $T^{ijk}_{pqr}$ to simplify the description~\cite{Hotta2009}
\begin{align}
T^{ijk}_{pqr}
=
\frac{1}{2}
\left(
\delta^{k}_{p}
\delta^{(i}_{r}\delta^{j)}_{q}
+
\delta^{k}_{q}
\delta^{(i}_{r}\delta^{j)}_{p}
-\delta^{k}_{r}
\delta^{(i}_{p}\delta^{j)}_{q}
\right)
\end{align}
with the notation 
$\delta^{(i}_{p}\delta^{j)}_{q}
=(\delta^{i}_{p}\delta^{j}_{q}
+\delta^{j}_{p}\delta^{i}_{q})$.
The term $P^{ij}$ originates from the boundary term of the CS action.
The polynomial $P_{ij}$ is called the Bardeen-Zumino polynomial~\cite{Bardeen:1984pm}.
%\textcolor{blue}{The polynomial $P_{ij}$ is called the Bardeen-Zumino polynomial \cite{Bardeen:1984pm}.}
In performing the integration by parts of the CS action, we imposed the boundary condition on the edge
$\delta g_{\mu\nu}|_{y=0}
=(\delta^{i}_{\mu}\delta^{j}_{\nu}+\delta^{j}_{\mu}\delta^{i}_{\nu}) \delta h_{ij}/2$.
Note that the terms $\Gamma^{y}_{ij}$ and $\Gamma^{i}_{j y}$ vanish because they do not contribute near the edge with the condition $f'(y)|_{y=0} = 0$.

In the third term of Eq.~\eqref{derivative}, we defined $T^{ij}_{\text{cov}}$ as~\cite{Bertlmann1997}
\begin{align}
T^{ij}_{\text{cov}}
=
T^{ij}_{\text{edge}}+P^{ij}.
%T^{ij}_{\text{cons}}+P^{ij}.
\end{align}
The covariant derivative of the above equation leads to
\begin{align}
\nabla_{i}T^{ij}_{\text{cov}}
=
\nabla_{i}T^{ij}_{\text{edge}}+\nabla_{i}P^{ij},
%\nabla_{i}T^{ij}_{\text{cons}}+\nabla_{i}P^{ij},
\end{align}
where $\nabla_{i}T^{ij}_{\text{cov}}$ satisfies the so-called \textit{covariant gravitational anomaly}~\cite{Bertlmann1997, Bertlmann2001} as follows:
\begin{align}
\nabla_{i}T^{ij}_{\text{cov}}
=
-\beta\bar{\epsilon}^{ij}\nabla_{i}R
\label{covequu}
\end{align}
with the scalar curvature $R$ in $(1+1)$-dimensions.
This equation has the covariant form with respect to the coordinate transformation.
Thus the term $P^{ij}$ converts $T^{ij}_{\text{cons}}$ to $T^{ij}_{\text{cov}}$.
This result means that the anomaly equation~\eqref{consistent} is modified by the flow of the conserved charges originating from the bulk (anomaly-inflow) and that the energy-momentum tensor on the edge becomes covariant~\cite{Stone2012, Jensen2013}.

%エネルギー保存則をリスペクトしたアノマリー方程式の導出（セクションはわけない
%（cotton tensorの役割がより明確になる））
%\textcolor{red}{TN:Write the energy conservation law in $2+1$d and write the anomalous conservation law at the edge. → then focus on the $1+1$d edge analysis.}
%Let us consider the general coordinate transformation $\delta_{\xi} g_{\mu\nu}
%=
%\nabla_{\mu}\xi_{\nu}+\nabla_{\nu}\xi_{\mu}$
%in the entire system.
%The variation is modified and leads to
%\begin{align}
%0=
%\delta_{\xi}W_{\text{tot}}
%=
%-\int d^3x\sqrt{-g}
%\nabla_{\mu}T^{\mu\nu}_{\text{bulk}}\xi_{\nu}
%-
%\int d^2x\sqrt{-h}
%\nabla_{i}T^{ij}_{\text{cov}}\xi_{j}
%\end{align}
In the above discussion, we obtained the anomaly equation~\eqref{covequu} from the diffeomorphism invariance of the generating functional $W_{\text{tot}}$.
In the following, we derive Eq.~\eqref{covequu} in terms of the conservation of energy and momentum to clarify the role of the bulk energy-momentum $T^{\mu\nu}_{\text{bulk}}$.
The conservation of the energy-momentum tensor in the total system is
\begin{align}
0=\nabla_{\mu}T^{\mu\nu}_{\text{tot}},
\label{totenergy}
\end{align}
where $T^{\mu\nu}_{\text{tot}}$ consists of two terms
\begin{align}
T^{\mu\nu}_{\text{tot}}
=
T^{\mu\nu}_{\text{edge}}
+T^{\mu\nu}_{\text{bulk}}
\end{align}
with $T^{\mu\nu}_{\text{edge}}$ being the $(2+1)$-dimensional version of the energy-momentum tensor on the edge.
A special solution of the differential equation~\eqref{totenergy} satisfies the following equations
\begin{align}
T^{\mu\nu}_{\text{edge}}
=
\begin{pmatrix}
T^{ij}(t,x)\delta(y)&O_{2\times1}\\
O_{1\times2}&0
\end{pmatrix}
,
\quad
T^{\mu\nu}_{\text{bulk}}
=
\begin{pmatrix}
O_{2\times2}&\nabla_{i}T^{ij}(t,x)\theta(-y)\\
\nabla_{i}T^{ij}(t,x)\theta(-y)&-\int dy\theta(-y)\nabla_{i}\nabla_{j}T^{ij}(t,x)
\end{pmatrix}.
%\begin{pmatrix}
%O_{2\times2}&-\nabla_{i}T^{ij}(t,x)\theta(y-y_{0})\\
%-\nabla_{i}T^{ij}(t,x)\theta(y-y_{0})&\int dy\theta(y-%y_{0})\nabla_{i}\nabla_{j}T^{ij}(t,x)
%\end{pmatrix}.
\end{align}
In the following, we will show that the tensor $T^{ij}(t, x)$ is identified as $T^{ij}_{\text{edge}}$.
By directly substituting these equations, we can verify that Eq.~\eqref{totenergy} is satisfied.

In the metric of Eq.~\eqref{edge} on the edge, 
$T^{\mu\nu}_{\text{bulk}}$
behaves as follows
\begin{align}
T^{2j}_{\text{bulk}}(t,x)
=
-2\beta\bar{\epsilon}^{ik2}\nabla_{i}R^{j}_{k}
=-\beta\bar{\epsilon}^{ij2}\nabla_{i}R,
\end{align}
where, in the second equality, we used the identity
$2R^{j}_{k}=\delta^{j}_{k}R(t,x)$, which holds in 
$(1+1)$-dimensions~\cite{MR0152974,Stone2012}.
Thus the tensor $T^{ij}$ has the same formula as~\eqref{covequu} as follows:
\begin{align}
\nabla_{i}T^{ij}
=
-\beta\bar{\epsilon}^{ij}\nabla_{i}R,
\end{align}
which satisfies the following identity
\begin{align}
\nabla_{i}\nabla_{j}T^{ij}
=0.
\end{align}
Therefore, the bulk energy-momentum tensor 
$T^{\mu\nu}_{\text{bulk}}$  
is related to the energy-momentum tensor of the edge mode via the anomaly equation.
%The general solution of the covariant anomaly equation is 
%\begin{align}
%T^{ij}
%=
%-\beta\bar{\epsilon}^{ij}R+C^{ij},
%\end{align}
%where $C^{ij}$ is a homogeneous solution of the anomaly equation, which will be fixed by imposing the boundary conditions in our setup.
Thus, the total energy-momentum tensor near the edge is only determined by $T^{ij}$.
In the next section, we discuss the energy flux analogous to the Hawking radiation.

%\textcolor{red}{TN:Write the general form of the stress tensor on a curved background.}

\section{Solution of anomaly equation on expanding edge}
In this section, we compute the energy flux of the energy-momentum tensor on the expanding edge in two ways.
First, we briefly review the setup of the QHS with the expanding edge.
Next, we solve the anomaly equation near the boundary of the expanding edge and derive for the energy flux.
Finally, we evaluate the energy flux using the general form of the Weyl anomaly.
Then, we show that the behavior at late time converges to a constant flux.

\subsection{Quantum Hall System with an expanding edge}
Here, we firstly introduce the coordinate structure of the QHS with an expanding edge and derive the matching conditions at the boundary between regions with different geometries. 
\begin{figure}[H]
 \centering
  \includegraphics[width=0.45\linewidth]{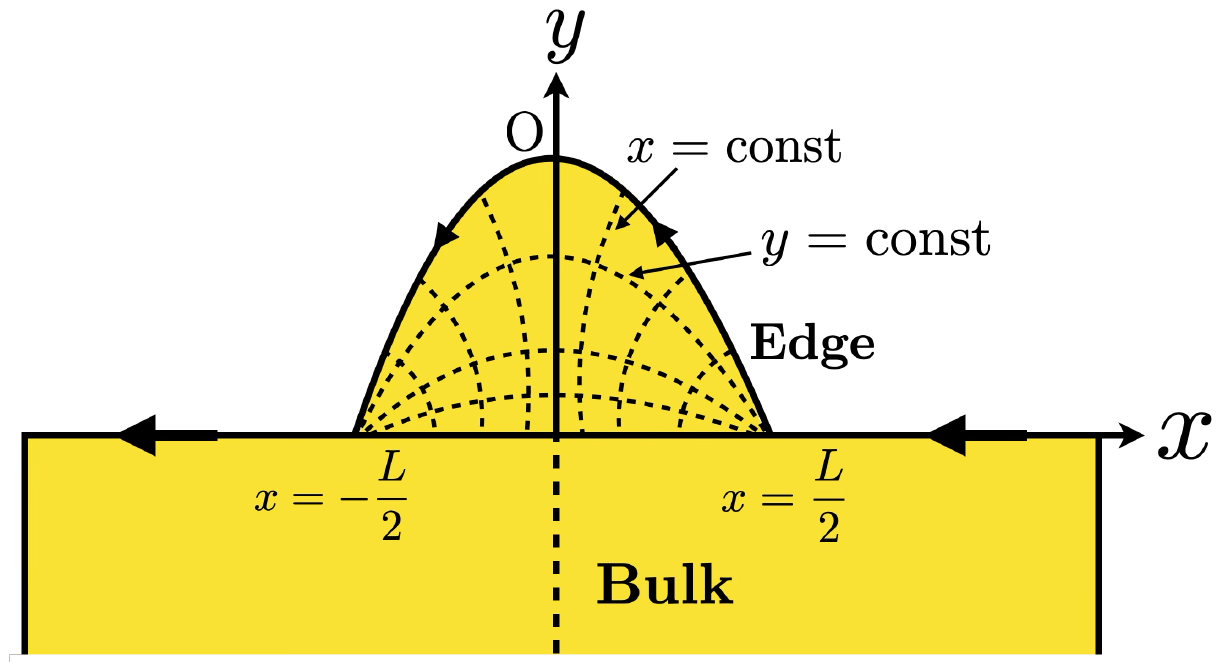}
   \caption
    {Schematic picture of the quantum Hall system with an expanding edge.
    The edge state exists at $y=0$ arrows, denoted by bold, while the region $y<0$ shows the bulk.
    The black dashed lines denote the lines where $x$ and $y$ coordinates are constant, respectively.
    }
\label{fig:qhs}
\end{figure}
\noindent
This setup serves as an analog model for curved spacetime, in particular a $(1+1)$-dimensional de Sitter universe~\cite{Hotta2022, Nambu2023, Yoshimoto2025}, and allows us to analyze the behavior of chiral edge states in the presence of time-dependent geometries.
A more detailed explanation is presented in Appendix B.
The metric of the expanding QHS in each region is given in null coordinates as (see Fig.~\ref{fig:qhs}):
\begin{align}
ds^2
=\omega^2(t,x)\eta_{ij}dx^{i}dx^{j}
&{
=
\left\{
\begin{array}{cr}
-dx^{+}_{\rm{I}}dx^{-}_{\rm{I}}
& ~~ ({\rm Region~I};\ x=x_{\rm{I}}>L/2)
\\ 
-e^{2\Theta(t)}dx^{+}dx^{-}
& ~~ ({\rm Region~II};\ |x|\leq L/2)
\\
-dx^{+}_{\rm{III}}dx^{-}_{\rm{III}}
& ~~ ({\rm Region~III};\ x=x_{\rm{III}}<-L/2)
\end{array}
\quad\right.
},
\label{metric}
\end{align}
where, in the second equality, the factor $\omega(t,x)$ is expressed as a single formula valid in all regions:
\begin{align}
\omega^2(t,x)
=
e^{2\Theta(t)}
\theta\left(\frac{L}{2}-x\right)
\theta\left(x+\frac{L}{2}\right)
\end{align}
with $\theta(x)$ being the step function.
In the last equality of Eq.~\eqref{metric}, we introduced the null coordinates $x^{\pm}$ as
$x^{\pm}:=t\pm x$ in the expanding region with the conformal factor $e^{\Theta(t)}$ and
$x^{\pm}_{i}:=t_{i}\pm x_{i}$
with $i=\rm{I},\ \rm{III}$.
We defined three regions; the region $\rm{I}$ and $\rm{III}$ describe the flat region, which is divided by the spatial coordinates 
$x=x_{\rm{I}}=L/2$ 
and 
$x=x_{\rm{III}}=-L/2$.
Here, the quantity $L$ characterizes the width of the region II when the edge is static.

At the boundary $x=x_{\rm{I}}=L/2$, the matching condition yields:
\begin{align}
x^{+}_{{\rm{I}}}[x^{+}]
=
\Phi[x^{+}-L/2]-\Phi[-L/2],
\label{coordinateI}
%\quad
%x^{-}_{\rm{I}}[x^{-}]=\Phi[x^{-}+L/2]-\Phi[-L/2]-L,
\end{align}
Similarly, at $x=x_{\rm{III}}=-L/2$, we obtain:
\begin{align}
x^{+}_{{\rm{III}}}[x^{+}]
=
\Phi[x^{+}+L/2]-\Phi[L/2],
\label{coordinateIII}
%\quad
%x^{-}_{\rm{III}}[x^{-}]=\Phi[x^{-}-L/2]-\Phi[L/2]+L,
\end{align}
where, in Appendix B, we present the detailed derivation of these matching conditions.
Note that to smoothly connect the coordinates between Region II and the outer flat regions (I and III), we introduce a function 
$\Phi[x]$ defined as:
\begin{align}
\Phi[x]:=\int_{0}^{x}dy e^{\Theta(y)}.
\label{phi}
\end{align}
This function plays the role of a connection between the expanding and static regions. 
Eliminating $x^{+}$ and combining Eqs.~\eqref{coordinateI} and~\eqref{coordinateIII}, we obtain the coordinate transformation between regions I and III:
\begin{align}
x^{+}_{\rm{I}}
&=
\Phi[-L+\Phi^{-1}[x^{+}_{\rm{III}}+\Phi[L/2]]]-\Phi[-L/2]
=:f[x^{+}_{\rm{III}}],
\label{transform}
%\\
%\quad
%x^{-}_{\rm{I}}
%&=
%\Phi[L+\Phi^{-1}[x^{+}_{\rm{III}}+\Phi[L/2]-L]]-\Phi[-L/2]-L
\end{align}
where
$\Phi^{-1}[x]$
is the inverse function of $\Phi[x]$.
This equation indicates that the coordinate of region I can be expressed by the coordinate of region III with the function 
$f[x^{+}_{\rm{III}}]$.

The conformal factor $e^{\Theta(t)}$ in Region II captures the expansion dynamics and is assumed to be of the de Sitter form~\cite{Hotta2022, Nambu2023, Yoshimoto2025}
%:{\color{red}$\omega(t ,x)=e^{\Theta(t)}$ is chosen as}
\begin{align}
%\omega(t, x)=
e^{\Theta(t)}
=\frac{1}{\cos(Ht)}
\end{align}
where the quantity $H^{-1}$ corresponds to the de Sitter curvature radius. 
This metric corresponds to a de Sitter universe in $(1+1)$-dimensions embedded within the edge of the QHS. 
The time domain is restricted to $|t|<\pi/2H$ due to the de Sitter horizon.
The functions $\Phi[x]$ and $\Phi^{-1}[x]$ are explicitly given by 
\begin{align}
\Phi[x]
=\frac{1}{2H}\ln\frac{1+\sin (Hx)}{1-\sin (Hx)},
\quad
\Phi^{-1}[x]
=\frac{1}{H}\sin^{-1}
\left[\tanh(Hx)
\right]. \label{eq:Phi}
\end{align}
In particular, the term $\Phi[-L/2]$ included in Eq.~\eqref{transform} leads to the antilogarithm condition
$1-\sin (HL/2)>0$, 
which satisfies
$HL<\pi$.
%\begin{align}
%\Phi\left[-\frac{L}{2}\right]
%=
%\frac{1}{2H}\ln\frac{1-\sin HL/2}{1+\sin HL/2}
%\end{align}

The metric in region II is described by the coordinates in regions I and III as
\begin{align}
dx^2_{\rm{II}}
&=
-\left[
\frac{1}{\cosh[H(x_{\rm{I}}-L/2)]}
\right]^2
dx^{+}_{\rm{I}}dx^{-}_{\rm{I}},
\label{metia}
\\
\quad
dx^2_{\rm{II}}
&=-\left[
\frac{1}{\cosh[H(x_{\rm{III}}+L/2)]}
\right]^2
dx^{+}_{\rm{III}}dx^{-}_{\rm{III}}.
\label{metiiib}
\end{align}
These line elements ensure that the conformal factor becomes unity at the boundaries 
$x=x_{\rm{I}}= L/2$ and 
$x=x_{\rm{III}}= -L/2$, and the spacetime can be extended smoothly across the boundaries. 
Furthermore, as shown in the Penrose diagram in Fig.~3 of Ref.~\cite{Yoshimoto2025}, the coordinate of region II is covered by the coordinates of regions I and III.
As in the previous study~\cite{Yoshimoto2025}, the expansion regions described by the coordinates of regions I and III are defined as region A (metric is Eq.~\eqref{metia}) and region B (metric is Eq.~\eqref{metiiib}).
In the next two subsections, we derive the energy flux in two ways.

\subsection
{Anomaly equation and Energy Flux}
Let us consider the anomaly equation 
\begin{align}
\nabla^{i}T_{ij}
=-\beta\bar{\epsilon}_{ij}\partial^{i}R
\label{coveqdd}
\end{align}
with $T_{ij}=h_{ik}h_{jl}T^{kl}$
and
$\bar{\epsilon}_{ij}
=h_{ik}h_{jl}\bar{\epsilon}^{kl}$
and derive the energy flux to solve it.
The previous study~\cite{Yoshimoto2025} considered the conditions for anomaly-free to ensure general coordinate transformation invariance on the edge and derived the energy flux formula in the outer flat region.
In this subsection, we show that a similar energy flux formula can be obtained by solving the anomaly equation Eq.~\eqref{coveqdd} near the boundary between the expanding and flat regions, rather than under conditions that are anomaly-free.

First, to solve the anomaly equation Eq.~\eqref{coveqdd}, we denote the components of the energy-momentum tensor by $T_{00}$.
This is achieved by imposing the following two conditions:
\begin{align}
T^{i}_{\ i}
&=
h^{ij}T_{ij}
=2\beta R,
\\
\quad
T_{--}
&=\frac{1}{4}
\left(T_{00}+T_{11}-2T_{01}\right)
=0.
\end{align}
The first condition of nonvanishing tensor trace is the Weyl anomaly~\cite{Bertlmann1997}, which leads to
\begin{align}
T_{00}-T_{11}=-2\beta \omega^2 R.
\label{weyl}
\end{align}
The second condition means that the energy current from the direction of $x^{+}$ component is zero due to the chirality of QHS and satisfies the following equation
\begin{align}
T_{00}+T_{11}-2T_{01}=0.
\label{chiral}
\end{align}
Combining Eqs.~\eqref{weyl} and~\eqref{chiral}, $T_{ij}$ is described only by $T_{00}$ component as
\begin{align}
T_{ij}
=
\begin{pmatrix}
T_{00}&&T_{00}+\beta\omega^2 R\\
T_{00}+\beta\omega^2 R&&T_{00}+2\beta\omega^2 R
\end{pmatrix}.
\end{align}
The components of the anomaly equations Eq.~\eqref{coveqdd} for $j=0$ and $j=1$ lead to the same equation as
\begin{align}
-(\partial_{0}-\partial_{1})T_{00}
-2\beta R\omega(\partial_{0}-\partial_{1})\omega
=-2\beta\omega^2\partial_{1} R.
\label{anomalyeq}
\end{align}

From now on, we solve for the anomaly equation Eq.~\eqref{anomalyeq} near the boundaries of regions I and III, respectively.
It is convenient to introduce the scalar curvature $R$ in our system as follows~\cite{Yoshimoto2025}
\begin{align}
R=
2H^2\theta\left(\frac{L}{2}-x_{\text{I}}\right)
\theta\left(\frac{L}{2}+x_{\text{III}}\right).
\end{align}
The curvature takes on non-zero values only in the expanding region. 
Let us first solve the Eq.~\eqref{anomalyeq} in the interval 
$\{x \in \mathbb{R}\ |\ L/2-\epsilon \leq x \leq L/2+\epsilon\}$
with positive infinitesimal $\epsilon\to+0$.
This interval represents the infinitesimal region of I and A containing $x=L/2$.
Substituting the scalar curvature
$R=2H^2\theta(L/2-x_{\text{I}})$
and the conformal factor
$\omega^2=\cosh^{-2}[H(x_{\rm{I}}-L/2)]$
in region A into Eq.~\eqref{anomalyeq} and integrating both sides of Eq.~\eqref{anomalyeq} from 
$x=L/2-\epsilon$ to 
$x=L/2+\epsilon$, we obtain the following equation
\begin{align}
\lim_{\epsilon\to+0}
\left[
T_{00}\left(t,\frac{L}{2}+\epsilon\right)
-T_{00}\left(t,\frac{L}{2}-\epsilon\right)
%-2\beta\int_{\frac{L}{2}-\epsilon}^{\frac{L}{2}+\epsilon} dx
%\left[\omega(\partial_{0}-\partial_{1})\omega R\right]
\right]
=
4\beta H^2,
%\lim_{\epsilon\to+0}
%\int_{\frac{L}{2}-\epsilon}^{\frac{L}{2}+\epsilon}
%dx \omega^2\delta\left(x-\frac{L}{2}\right)
\label{regionsia}
\end{align}
where we evaluated the integral on the left-hand side as
\begin{align}
-\partial_{0}
\int_{\frac{L}{2}-\epsilon}^{\frac{L}{2}+\epsilon}
dx T_{00}
\xrightarrow{\epsilon\to+0}0,
\quad
-2\beta
\int_{{L}/{2}-\epsilon}^{{L}/{2}+\epsilon}dx
\omega(\partial_{0}-\partial_{1})\omega R
=
-4\beta H^2\int_{{L}/{2}-\epsilon}^{{L}/{2}}dx\omega(\partial_{0}-\partial_{1})\omega 
\xrightarrow{\epsilon\to+0}0.
\end{align}
Since the derivative of the step function results in a delta function, the right-hand side is evaluated as
\begin{align}
-4\beta H^2 
\lim_{\epsilon\to+0}
\int_{L/2-\epsilon}^{L/2+\epsilon}
\omega^2\partial_{1}
\theta\left(\frac{L}{2}-x\right)
=
4\beta H^2 
\lim_{\epsilon\to+0}
\int_{\frac{L}{2}-\epsilon}^{\frac{L}{2}+\epsilon}
dx \omega^2\delta\left(x-\frac{L}{2}\right)
=
4\beta H^2 
\end{align}
with $\omega^2|_{x=L/2} =1$.
Similarly, in the interval
$\{x \in \mathbb{R}\ |\ -L/2-\epsilon \leq x \leq -L/2+\epsilon\}$
with positive infinitesimal $\epsilon\to+0$,
the integral of the anomaly equation Eq.~\eqref{anomalyeq} reads
%\begin{align}
%-\partial_{0}
%\int_{-\frac{L}{2}-\epsilon}^{-\frac{L}{2}+\epsilon}
%dx T_{00}
%+T_{00}\left(t,-\frac{L}{2}+\epsilon\right)
%-T_{00}\left(t,-\frac{L}{2}-\epsilon\right)
%-2\beta\int_{-\frac{L}{2}-\epsilon}^{-\frac{L}%{2}+\epsilon} dx
%\left[\omega(\partial_{0}-\partial_{1})\omega %R\right]
%=
%-4\beta H^2
%\int_{-\frac{L}{2}-\epsilon}^{-\frac{L}{2}+\epsilon}
%dx \omega^2\delta\left(x+\frac{L}{2}\right)
%\end{align}
\begin{align}
\lim_{\epsilon\to+0}
\left[
T_{00}\left(t, -\frac{L}{2}+\epsilon\right)
-T_{00}\left(t, -\frac{L}{2}-\epsilon\right)
\right]
=-4\beta H^2
\label{regionsiiib}
\end{align}
with the scalar curvature 
$R=2H^2\theta(x_{\text{III}}+L/2)$
and the conformal factor
$\omega^2=\cosh^{-2}[H(x_{\rm{III}}+L/2)]$.
in region B.

We focus on the energy flux of the left-moving mode $T_{++}$.
This component is related to $T_{00}$ as follows:
\begin{align}
T_{++}
=\frac{1}{4}
\left(T_{00}+T_{11}+2T_{01}\right)
=T_{00}+\beta\omega^2 R.
\label{pp00}
\end{align}
Using the above equation Eq.~\eqref{pp00}, $T_{++}$ component of each of the regions I and A is defined as
\begin{align}
T^{\text{I}}_{++}
:=
\lim_{\epsilon\to+0}
T_{00}\left[t, \frac{L}{2}+\epsilon\right],
\quad
T^{\text{A}}_{++}
:=
\lim_{\epsilon\to+0}
T_{00}\left[t, \frac{L}{2}-\epsilon\right]+2\beta H^2.
\end{align}
Similarly, in regions III and B respectively, $T_{++}$ is written as follows:
\begin{align}
T^{\text{III}}_{++}
:=
\lim_{\epsilon\to+0}
T_{00}\left[t, -\frac{L}{2}-\epsilon\right],
\quad
T^{\text{B}}_{++}
:=
\lim_{\epsilon\to+0}
T_{00}\left[t, -\frac{L}{2}+\epsilon\right]+2\beta H^2
\end{align}
Thus, Eqs.~\eqref{regionsia} and~\eqref{regionsiiib} yield the following equations
\begin{align}
T^{\text{A}}_{++}=-2\beta H^2+T^{\text{I}}_{++},
\quad
T^{\text{III}}_{++}=2\beta H^2+T^{\text{B}}_{++}.
\end{align}
We impose the boundary condition 
$T^{\text{I}}_{++}=0$, which means that there is no energy flux from the beginning (in-vacuum condition).
We also relate to 
$T^{\text{A}}_{++}$ and
$T^{\text{B}}_{++}$.
This is achieved by doing the general coordinate transformation in the overlapped region 
$A\cap B$~\cite{Yoshimoto2025} as follows:
\begin{align}
T^{\text{B}}_{++}
=
\left(
\frac{\partial x^{+}_{\text{I}}}
{\partial x^{+}_{\text{III}}}
\right)^2
T^{\text{A}}_{++},
\end{align}
where the coordinate transformation is performed by Eq.~\eqref{transform}.
The function $\partial x^{+}_{\text{I}}/\partial x^{+}_{\text{III}}$ is explicitly given by
\begin{align}
\frac{\partial x^{+}_{\text{I}}}
{\partial x^{+}_{\text{III}}}
=
\frac{1}{\cos(HL)+2\sin(HL/2)
\left(
\sinh[H x^{+}_{\text{III}}]
+\sin(HL/2)\cosh[H x^{+}_{\text{III}}]
\right)}.
\end{align}
Thus, the energy flux of the left-moving mode  region III
$T^{\text{III}}_{++}$
is
\begin{align}
T^{\text{III}}_{++}
=
2\beta H^2 +T^{\text{B}}_{++}
=
2\beta H^2
\left[
1-\left(
\frac{\partial x^{+}_{\text{I}}}
{\partial x^{+}_{\text{III}}}
\right)^2
\right].
\label{ppcomponent}
\end{align}
This formula approaches 
$2\beta H^2=c_{-}H^2/48\pi$ 
in the late time limit
$x^{\text{III}}_{+}\to\infty$
in region III.
%\begin{figure}[H]
% \centering
%  \includegraphics[width=0.5\linewidth]{result.pdf}
%   \caption
%    {The function $T^{\rm{III}}_{++}$ as a function of $Hx^{+}_{\text{III}}.$}
%\label{fig:tpp}
%\end{figure}
%In particular, assuming the thermal equilibrium with the temperature $T$ at the late time, the chiral edge mode carry the energy flux 
%$I={\pi c_{-}T^2}/{12}$~\cite{Stone2012, Cappelli2002, Kitaev2006}, which reproduces the Gibbons-Hawking temperature $T_{\text{GH}}=H/2\pi$.
Using the Gibbons-Hawking temperature $T_{\text{GH}}=H/2\pi$~\cite{GH1977} for de Sitter space, we obtain exactly the energy flux carried by the chiral edge mode in thermal equilibrium $I={\pi c_{-}T^2}/{12}$ \cite{Stone2012, Cappelli2002, Kitaev2006} at the temperature $T=T_{\text{GH}}$.
They are consistent with the result obtained in the previous study~\cite{Yoshimoto2025}.
In the next subsection, we consider the general form of the Weyl anomaly and energy-momentum tensor discussed in Ref.~\cite{Maldacena2018}, and then derive the thermal flux formula without using the anomaly equation.

\subsection{Weyl anomaly and Energy Flux}
%\textcolor{blue}{TN: edit this section}
Using the transformation of the partition function under the Weyl transformations~\cite{Polyakov:1981rd,Maldacena2018} in conformal field theory without gravitational anomalies, we can derive the transformation of the energy-momentum tensor in $(1+1)$-dimensional gapless relativistic theory under the Weyl transformations. % by the general form of the conformal transformation
When we perform the Weyl transformation as $h_{ij}=e^{2\Omega}\hat{h}_{ij}$, the stress tensor is transformed as
\begin{align}
t_{ij}
=
\hat{t}_{ij}
-\frac{c}{12\pi}
\left[
\partial_{i}\Omega\partial_{j}\Omega
-\hat{{\nabla}}_{j}\hat{{\nabla}}_{i}\Omega
-\frac{1}{2}\hat{g}_{ij}(\hat{{\nabla}}\Omega)^2
+\hat{g}_{ij}\hat{{\nabla}}^2\Omega
\right],
\label{maldacena}
\end{align}
where the tensor $\hat{t}_{ij}$ is the energy-momentum tensor before performing the Weyl transformation, i.e., the expectation value of the stress tensor under the metric $\hat{h}_{ij}$.
The second term stems from the conformal transformation with 
$\Omega$
%and $c$ 
being the general conformal factor.
%and central charge. 
$c$ is the total central charge which is the sum of the central charges of chiral and anti-chiral modes.
For chiral gapless theories, the total central charge is equal to the chiral central charge $c_{-}$ because we only have chiral modes.
In our setup, $\hat{t}_{ij}$ is defined in the outer flat regions \eqref{metric}.
Since $t_{ij}$ is defined on a flat spacetime background, there is no contribution from the Weyl anomaly: 
$\hat{t}^{i}_{\ i}
=\hat{h}^{ij}\hat{t}_{ij}=0$.
The trace of the second term originated from the conformal transformation is proportional to the scalar curvature and contributes to the Weyl anomaly
\begin{align}
t^{i}_{\ i}
=
h^{ij}t_{ij}
=\frac{c}{24\pi}R. \label{eq:STWeyl}
\end{align}
Using Eq.~\eqref{maldacena}, the $t_{++}$ component in the null coordinates becomes 
\begin{align}
t_{++}
=
\frac{1}{4}\left[t_{00}+t_{11}+t_{01}+t_{10}\right]
=
\hat{t}_{++}
+
\frac{c}{12\pi}
\left(\partial^2_{+}\Omega-(\partial_{+}\Omega)^2\right),
\label{weylpp}
\end{align}
where we used $\partial_{0}+\partial_{1}=2\partial_{+}$.
$t_{++}$ on the left-hand side is defined by the curved spacetime metric, whereas, on the right-hand side, $\hat{t}_{++}$ is defined by the flat spacetime metric.

\begin{figure}[H]
 \centering
  \includegraphics[width=0.6\linewidth]{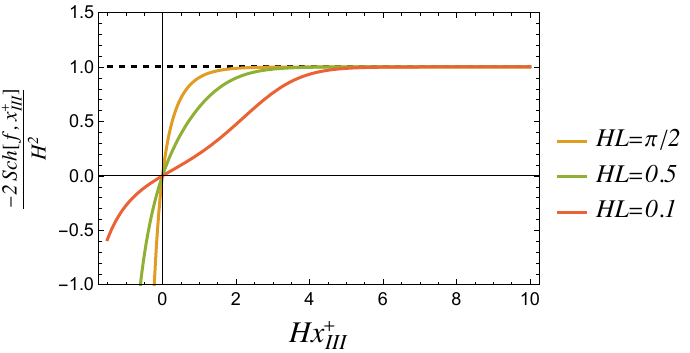}
    \caption
     {The function 
     $-2\text{Sch}[f, x^{+}_{\rm{III}}]/H^2$ as a function of $Hx^{+}_{\text{III}}$ for 
     $HL=0.1, 0.5, \pi/2$.}
\label{fig:scwaltz}
\end{figure}

In the following, we see that $\hat{t}_{++}$ in region III corresponds to the energy flux originating from the anomalous term in the second term of Eq.~\eqref{weylpp}.
We consider the following coordinate transformation as
\begin{align}
ds^2
=
-dx^{+}_{\rm{I}}dx^{-}_{\rm{I}}
=
-\left(\frac{dx^{+}_{\rm{I}}}{dx^{+}_{\rm{III}}}\right)\left(\frac{dx^{-}_{\rm{I}}}{dx^{-}_{\rm{III}}}\right)dx^{+}_{\rm{III}}dx^{-}_{\rm{III}}
=:-e^{2\Omega}dx^{+}_{\rm{III}}dx^{-}_{\rm{III}},
%-e^{2\omega[x^{+}_{\rm{I}}, x^{-}_{\rm{I}}]}dx^{+}_{\rm{III}}dx^{-}_{\rm{III}}
\end{align}
where, in the second equality, we changed the coordinate from the region I to III.
In the last equality, we introduced the general conformal factor $\Omega$, which is explicitly written as
\begin{align}
\Omega
=
\frac{1}{2}\log\left(\frac{dx^{+}_{\rm{I}}}{dx^{+}_{\rm{III}}}\right)+\frac{1}{2}\log\left(\frac{dx^{-}_{\rm{I}}}{dx^{-}_{\rm{III}}}\right).
\end{align}
Thanks to Eq.~\eqref{weylpp}, the energy flux in region III 
$t^{\rm{III}}_{++}$ with the line element
$ds^2=-e^{2\Omega}dx^{+}_{\rm{III}}dx^{-}_{\rm{III}}$ is obtained as
\begin{align}
t^{\rm{III}}_{++}
=
\hat{t}^{\rm{III}}_{++}
+
\frac{c}{12\pi}
\left[
\Omega^{\prime\prime}-(\Omega^{\prime})^2
\right]
=
\hat{t}^{\rm{III}}_{++}
+
\frac{c}{24\pi}\text{Sch}[f, x^{+}_{\rm{III}}],
%\\
%\quad
%t_{\rm{III-}\rm{III-}}
%&=
%\hat{t}_{\rm{III-}\rm{III-}}
%+\frac{c}{12\pi}
%\left[
%\partial^2_{\rm{III-}}\omega-(\partial_{\rm{III-}}\omega)^2
%\right]
%\\
%\quad
%t_{\rm{III+}\rm{III-}}
%&=
%t_{\rm{III-}\rm{III+}}
%=
%\hat{t}_{\rm{III+}\rm{III-}}
%-\frac{c}{12\pi}\partial_{\rm{III+}}\partial_{\rm{III}-}\omega
\end{align}
where, in the second equality, 
$\ ^{\prime}$ denotes the derivative with respect to $x^{+}_{\rm{III}}$, respectively.
%where the symbol $\partial_{\rm{III}+}$ means the derivative with respect to $+$ component of the region III.
%%%%%%%%%%%%%%%%%%%%%%%%%%%%%%%%%%%%%%%%%%%%%
\if0
\begin{align}
\partial_{\rm{III}+}\Omega
=
\frac{1}{2}\partial_{\rm{III}+}\log\left(\frac{df[x^{+}_{\rm{III}}]}{dx^{+}_{\rm{III}}}\right)
=
\frac{f^{\prime\prime}}{2f^{\prime}},
\quad
\partial^2_{\rm{III}+}\Omega
=
\frac{1}{2}
\left[\frac{f^{\prime\prime\prime}}{f^{\prime}}
-\left(\frac{f^{\prime\prime}}{f^{\prime}}\right)^2
\right],
\end{align}
\begin{align}
\frac{c}{12\pi}
\left[
\partial^2_{\rm{III}+}\Omega-(\partial_{\rm{III}+}\Omega)^2
\right]
=
\frac{c}{24\pi}
\left[
\frac{f^{\prime\prime\prime}}{f^{\prime}}
-\frac{3}{2}\left(\frac{f^{\prime\prime}}{f^{\prime}}\right)^2
\right]
=
\frac{c }{24\pi}\text{Sch}[f, x^{+}_{\rm{III}}],
\end{align}
\fi
%%%%%%%%%%%%%%%%%%%%%%%%%%%%%%%%%%%%%%%%%%%%%
In the last equality, we defined the Schwarzian derivative
$\text{Sch}[f, x^{+}_{\rm{III}}]:={f^{\prime\prime\prime}}/{f^{\prime}}
-{3}/{2}\left({f^{\prime\prime}}/{f^{\prime}}\right)^2.$
Note that 
$t^{\rm{III}}_{++}$
is connected to 
$t^{\rm{I}}_{++}$ in region I by the coordinate transformation
$x^{+}_{\rm{I}}=f[x^{+}_{\rm{III}}]$
as follows:
\begin{align}
t^{\rm{III}}_{++}
=
\left(
\frac{\partial x^{+}_{\rm{I}}}{\partial x^{+}_{\rm{III}}}
\right)^2
t^{\rm{I}}_{++}.
\end{align}
Imposing the boundary condition
$t^{\rm{I}}_{++}=0$,
which means that there is no energy flux initially in region I, the energy flux in the outer region
$\hat{t}^{\rm{III}}_{++}$ 
becomes
\begin{align}
\hat{t}^{\rm{III}}_{++}
=
\frac{c H^2}{48\pi}
\left[
-\frac{2\text{Sch}[f, x^{+}_{\rm{III}}]}{H^2}
\right].
\label{weylppcomponent}
\end{align}
Fig.~\ref{fig:scwaltz} shows the behavior of the function 
$-2\text{Sch}[f, x^{+}_{\rm{III}}]/H^2$ with respect to $HL=0.1, 0.5, \pi/2$.
All curves in Fig.~\ref{fig:scwaltz} saturate at unity at the late time
$Hx^{+}_{\rm{III}}\to\infty$.
Thus, at the late time 
$Hx^{+}_{\rm{III}}\to\infty$, 
the energy flux
$\hat{t}^{\rm{III}}_{++}$ approaches the constant quantity $c H^2/48\pi$ represented by the black dashed line.
This is also consistent with the results of Eq.~\eqref{ppcomponent} and the previous study~\cite{Yoshimoto2025}.
Here, we provided an alternative derivation using only Weyl anomaly.
%However, we would like to emphasize that we can evaluate the expected energy flux without using the anomaly equation~\eqref{anomalyeq}.
%{\color{red}In fact, it can be shown analytically that the following equation holds
%\begin{align}
%-\frac{2\text{Sch}[f, x^{+}_{\rm{III}}]}{H^2}
%=1-(f^{\prime}[x^{+}_{\rm{III}}])^2,
%\end{align}
%so that the energy flux formula Eq.~\eqref{ppcomponent} is equivalent to Eq.~\eqref{weylppcomponent}.}

The formula~\eqref{maldacena} for the stress tensor is for non-chiral theories without gravitational anomalies.
However, we show that a special solution satisfying the anomaly equation~\eqref{anomalyeq} can be expressed using the components of the energy momentum tensor~\eqref{maldacena} for theories without gravitational anomalies.

First, we define a tensor 
$\mathcal{T}_{ij}$ 
obtained by normalizing $t_{ij}$ with the central charge $c$ as follows
\begin{align}
\mathcal{T}_{ij} = \frac{1}{c}t_{ij}.
\end{align}
Since the divergence of $t_{ij}$ is anomaly free $\nabla^{i}t_{ij}=0$, the $j=+$ component leads to
\begin{align}
0&=\nabla^{i}\mathcal{T}_{i+}
=\nabla^{+}\mathcal{T}_{++}+\nabla^{-}\mathcal{T}_{-+}.
%0&=\nabla^{i}t_{i-}=\nabla^{+}t_{+-}+\nabla^{-}t_{--},
\label{anofree}
\end{align}
Moreover %, realizing that $c = c_{\text{R}}+c_{\text{L}} =2c_{\text{R}}=2c_{\text{L}}= c_{-}$ in chiral theories, 
due to the Weyl anomaly 
$h^{ij}\mathcal{T}_{ij}= R/24\pi$ in the null coordinate, 
$t_{+-}$ and $t_{-+}$ components yield the following equation
\begin{align}
\mathcal{T}_{+-}
=\mathcal{T}_{-+}
=-\omega^2 \frac{1}{96\pi} R.
\end{align}
Let us introduce the following tensor
$\tilde{t}_{ij}$
\begin{align}
\tilde{t}_{ij}
= c_{-}
\begin{pmatrix}
\mathcal{T}_{++}&&\frac{1}{2}\mathcal{T}_{+-}\\
\frac{1}{2}\mathcal{T}_{-+}&&0
\end{pmatrix},
\end{align}
and we see that this tensor satisfies the anomaly equation Eq.~\eqref{anomalyeq} in the null coordinate.
The divergence for $j=+$ component leads to
\begin{align}
\nabla^{i}\tilde{t}_{i+}
=
%\nabla^{+}\tilde{t}_{+-}+\nabla^{-}\tilde{t}_{--}=
c_{-}\nabla^{+}{\mathcal{T}}_{++}+\frac{1}{2}c_-\nabla^{-}{\mathcal{T}}_{+-}
=
-\frac{1}{2}c_{-}\nabla^{-}\mathcal{T}_{+-},
\end{align}
where, in the last equality, we used Eq.~\eqref{anofree}.
Thus, 
$\nabla^{i}\tilde{t}_{i+}$ 
satisfies the following anomaly equation
\begin{align}
\nabla^{i}\tilde{t}_{i+}
=
-\frac{1}{2}\nabla^{-}(-\omega^2\beta R)
=-\beta\nabla^{-}(\bar{\epsilon}_{-+} R)
=-\beta\bar{\epsilon}_{-+} \nabla^{-}R,
\end{align}
where we used the covariant totally anti-symmetric tensor in the null coordinate 
$\bar{\epsilon}_{-+}$.
In Appendix A, we present the formulas such as $\bar{\epsilon}_{-+}$.
Similarly for $j=-$ component , we can derive the following anomaly equation 
\begin{align}
\nabla^{i}\tilde{t}_{i-}
=
\frac{1}{2}c_{-}\nabla^{+}{\mathcal{T}}_{+-}
%\frac{1}{2}\nabla^{+}(-\omega^2\beta R)
=-\beta\nabla^{+}(\bar{\epsilon}_{+-} R)
=-\beta\bar{\epsilon}_{+-} \nabla^{+}R.
\end{align}
Therefore, the tensor $\tilde{t}_{ij}$ is a solution of the anomaly equation~\eqref{anomalyeq} in the null coordinate.

Since the $\tilde{t}_{++} = c_-\mathcal{T}_{++}$ components are the same as those of the non-chiral theories except that $c$ is replaced by $c_-$, the emission of the Hawking radiation also takes the same form as that in non-chiral theories in Eq.~\eqref{weylppcomponent}: 
\begin{align}
\hat{t}^{\rm{III}}_{++}
=
\frac{c_- H^2}{48\pi}
\left[
-\frac{2\text{Sch}[f, x^{+}_{\rm{III}}]}{H^2}
\right].
\label{weylppcomponentc}
\end{align}
Using the definition of $\Phi[x]$ in~\eqref{eq:Phi}, this expression precisely agrees with ~\eqref{ppcomponent}.
The details of the derivation are presented in Appendix C.

So far, we have focused on the energy flux originating from the gravitational anomaly's anomalous current with $(1+1)$-dimensional expanding edges.
In the next section, we investigate the behavior of the bulk energy-momentum tensor in Eq.~\eqref{bulk} near the expanding edge region.

\section{Behavior of the bulk Energy-Momentum tensor}
In this section, we discuss the behavior of the bulk energy-momentum tensor~\eqref{bulk} that is given by the  $(2+1) d$ bulk metric.
The purpose of this section is to quantify how the bulk compensates the covariant gravitational anomaly that appears on the expanding edge. 
Translational invariance along the edge ($x$-direction) is assumed so that only the normal coordinate $y$ captures the crossover from the flat interior to the curved edge region:
\begin{align}
ds^2
=g_{\mu\nu}dx^{\mu}dx^{\nu}
=a^2(t,y)\eta_{ij}dx^{i}dx^{j}+dy^2,
\quad
a^2(t,y)
=\big(\omega(t)^2-1\big)f(y)+1,
\quad  
\omega (t) = \frac{1}{\cos(Ht)}.
\end{align}
We consider the situation where only the near edge region is expanding smoothly connected to the flat spacetime.
%In particular, we consider
To model a gradual edge–bulk crossover we first adopt the case where the spatial profile function $f(y)$ is of a Gaussian function case,
$f(y)=e^{-(y/y_{0})^n}$, 
and a function 
$f(y)=1/\{e^{n(y/y_{0}-1)}+1\}$
that takes the same form as the Fermi-Dirac distribution.
Note that the quantity
$y_{0}$
corresponds to %the boundary of the bulk 
the transition line between flat and curved regions.
The parameter $n$ controls the behavior of the function $f(y)$ near the bulk and edge boundaries.

The system is invariant under the combination of the reflection $R: (t,x,y) \to (t, -x ,y)$ and the sign flip $c_- \to -c_-$ \footnote{In quantum hall setups, we have electrons on a 2d plane under an external magnetic field.
In this setup, this transformation corresponds to the combination of a 180-degree rotation about the y-axis and the flip of the magnetic field $\bm{B} \to -\bm{B}$.}.
This transformation acts on the bulk stress tensor as 
\begin{align}
T^{\text{bulk}}_{\mu\nu} = 
\begin{pmatrix}
T_{00}^{\text{bulk}} & T_{01}^{\text{bulk}} & T_{02}^{\text{bulk}} \\
T_{10}^{\text{bulk}} & T_{11}^{\text{bulk}} & T_{12}^{\text{bulk}} \\
T_{20}^{\text{bulk}} & T_{21}^{\text{bulk}} & T_{22}^{\text{bulk}} \\
\end{pmatrix}
\to  
\begin{pmatrix}
-T_{00}^{\text{bulk}} & T_{01}^{\text{bulk}} & -T_{02}^{\text{bulk}} \\
T_{10}^{\text{bulk}} & -T_{11}^{\text{bulk}} & T_{12}^{\text{bulk}} \\
-T_{20}^{\text{bulk}} & T_{21}^{\text{bulk}} & -T_{22}^{\text{bulk}} \\
\end{pmatrix}
\end{align}
Therefore, only two components $T_{01}^{\text{bulk}}, T_{12}^{\text{bulk}}$ remain when the metric has reflection symmetry along $x$-direction, and we can focus on them.
%[Perhaps we don't need this sentence here because we didn't think about the $x$ dependence in this section.] Moreover, we focus on the parameter region where 
%$|x|<L/2$, $|t|<\pi/2H$ are satisfied, which denotes the expanding edge region. 
%We also focus on the parameter regime $HL<\pi/2$ where the signal can reach from the region I to region III in \eqref{metric}~\cite{Hotta2022}. 
%the other side of the expanding region \cite{Hotta2022}.
%The non-zero components of the bulk energy-momentum tensor are 
$T^{\text{bulk}}_{01}$ and $T^{\text{bulk}}_{12}$ correspond to energy currents parallel to the edge and the shear stress, respectively.
Their detailed expressions are given in Appendix D.

\subsection{Gaussian function case}
Firstly, we begin by considering a specific form of the spatial profile function $f(y)$ defined as:
\begin{align}
f(y)
=
e^{-(y/y_{0})^n}.
\label{gauss}
\end{align}
To satisfy the boundary conditions
$f(y)|_{y=0}=1$ and $f^{\prime}(y)|_{y=0}=0$,
we impose the constraint $n\geq2$.
These conditions ensure a smooth matching between the edge and the bulk region at $y=0$.
\begin{figure}[H]
  \centering
  \includegraphics[width=0.58\linewidth]{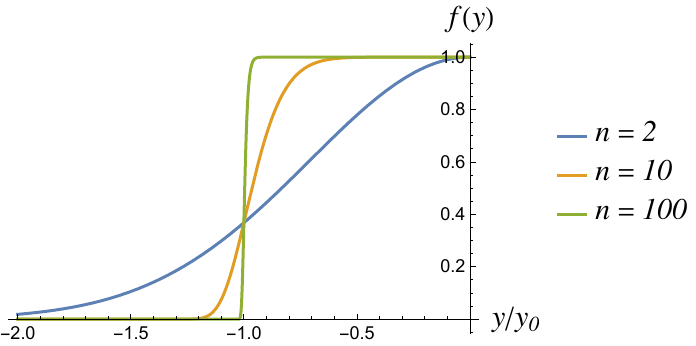}
\caption
{
 The behavior of the function 
 $f(y)=e^{-(y/y_{0})^n}$
 as a function of $y/y_{0}$ for $n=2, 10, 100$.
}
  \label{fig:gauss}
\end{figure}
\noindent
Fig.~\ref{fig:gauss} illustrates the behavior of the function $f(y)$ for different values of
$n=2, 10, 100$
in the bulk region $y<y_{0}$.
As $n$ increases, the function suddenly changes near $y=y_{0}$, which is similar to the Fermi-Dirac distribution function.
Although both forms fulfill the boundary conditions
$f(y)|_{y=0}=1$ and $f^{\prime}(y)|_{y=0}=0$ 
in the limit 
$n\to\infty$, for finite 
$n$, 
the Fermi-Dirac type does not satisfy the boundary conditions at 
$y=0$, 
resulting in distinct energy-momentum behavior near the edge.
In the next subsection, we also discuss $f(y)$ as the Fermi-Dirac distribution function case.
\begin{figure}[H]
  \centering
  \includegraphics[width=0.6\linewidth]{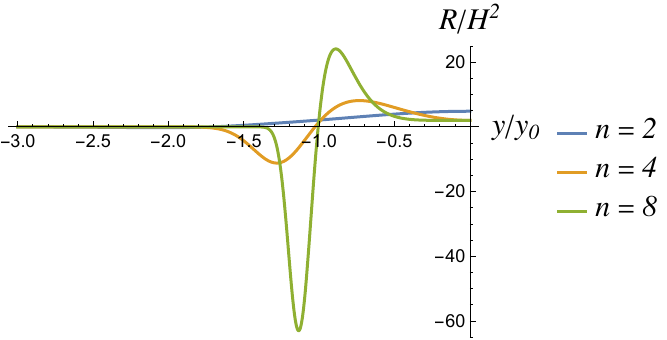}
\caption
{
 The behavior of the scalar curvature $R/H^2$ as a function of $y/y_{0}$ for $n=2, 4, 8$.
  We adopted $c_{-}=1$, and $Ht=1$.
 }
  \label{fig:curvegauss}
\end{figure}
The non-trivial behavior of
$f(y)$
leads to the nonzero scalar curvature 
$R$ near $y=y_{0}$ in the bulk.
Fig.~\ref{fig:curvegauss} shows the scalar curvature as a function of $y/y_{0}$ for $n=2, 4, 8$.
For $n=2$, the second derivative $f^{\prime\prime}$ 
at $y=0$ is nonzero, resulting in a discontinuity in curvature at the boundary.
In contrast $n>3$, the curvature remains smooth at $y=0$.
These results indicate the existence of a bulk region with nonzero curvature, which implies the possibility of induced energy-momentum flow within the bulk.
%{\color{red}(What means the scalar curvature takes both positive and negative values?)}
\begin{figure}[H]
  \centering
  \begin{minipage}[b]{0.49\linewidth}
  \includegraphics[width=1\linewidth]{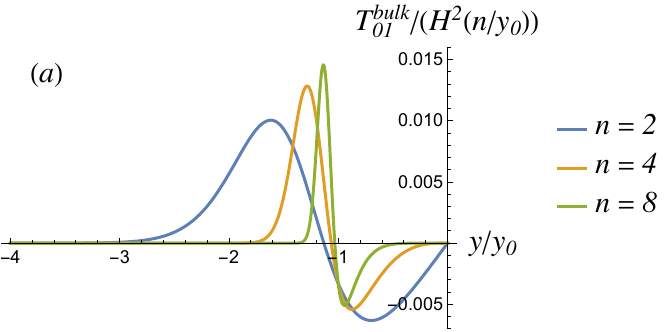}
  \end{minipage}\hspace{0.2cm}
  \begin{minipage}[b]{0.49\linewidth}
  \includegraphics[width=1\linewidth]{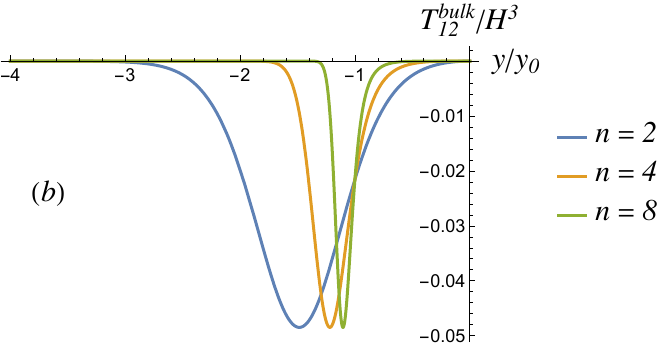}
  \end{minipage}
  \caption
{
 The behavior of the energy-momentum tensor in the bulk as a function of $y/y_{0}$ for $n=2, 4, 8$.
 Left and right panels correspond to 
 $T^{\text{bulk}}_{01}/(H^2(n/y_{0}))$ and 
 $T^{\text{bulk}}_{12}/H^3$.
  We adopted $c_{-}=1$, and $Ht=1$.
 }
  \label{fig:gauss0112}
\end{figure}
The existence of nonzero scalar curvature leads to the non-trivial behavior of the bulk energy-momentum tensor.
The panels (a) and (b) in Figs.~\ref{fig:gauss0112}
show the behavior of 
$T^{\text{bulk}}_{01}$
and
$T^{\text{bulk}}_{12}$, respectively.
Each curve in the panels presents the 
$n=2, 4, 8$.
In the panel (a), the curves exhibit oscillatory behavior and take both positive and negative values.
On the other hand, in the panel (b), each curve of $T^{\text{bulk}}_{12}$ is consistently negative throughout the region.
As $n$ increases for both the panel (a) and (b), the spatial spread of both components becomes more localized near $y=y_{0}$.
This suggests that the energy currents are confined to a narrow layer of thickness $\sim y_{0}/n$, centered around the interface between the flat and curved regions. 
These findings confirm the presence of non-trivial energy-momentum currents in the bulk, induced by the curved geometry near the expanding edge. 
This provides further evidence for the bulk–edge interplay governed by the anomaly-inflow mechanism.

\subsection{Fermi-Dirac distribution function case}
Next, we examine an alternative spatial profile function 
$f(y)$, defined by the Fermi-Dirac distribution
\begin{align}
f(y)
=
\frac{1}{e^{n(y/y_{0}-1)}+1}
\label{Dirac}
\end{align}
This function also describes a smooth transition between the edge and bulk regions.
\begin{figure}[H]
  \centering
  \includegraphics[width=0.58\linewidth]{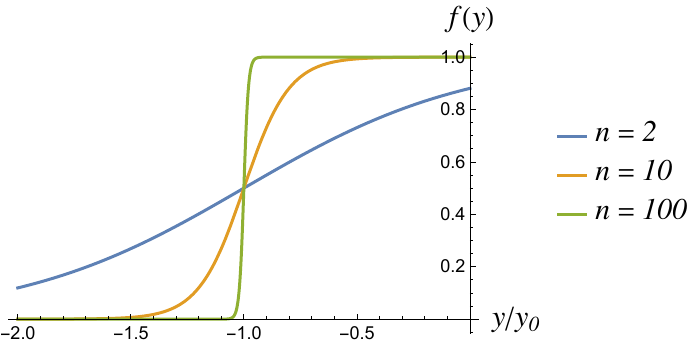}
\caption
{
 The behavior of the function 
 $f(y)=1/\{e^{n(y/y_{0}-1)}+1\}$
 as a function of $y/y_{0}$ for $n=2, 10, 100$.
 %We adopted $y_{0}=-1$.
 }
  \label{fig:dirac}
\end{figure}
Fig.~\ref{fig:dirac} shows the behavior of 
$f(y)$ for 
$n=2,10,100$
$y/y_{0}$.
Similar to the Gaussian profile discussed in the previous subsection, increasing $n$ sharpens the transition at 
$y=y_{0}$, effectively localizing the edge region. 
However, a key difference arises at the boundary 
$y=0$: this Fermi-Dirac type does not satisfy the conditions 
$f(y)|_{y=0}=1$ and 
$f^{\prime}(y)|_{y=0}=0$
for finite $n$. 
As a result, this deviation affects the energy-momentum tensor at the edge.
\begin{figure}[H]
  \centering
  \includegraphics[width=0.6\linewidth]{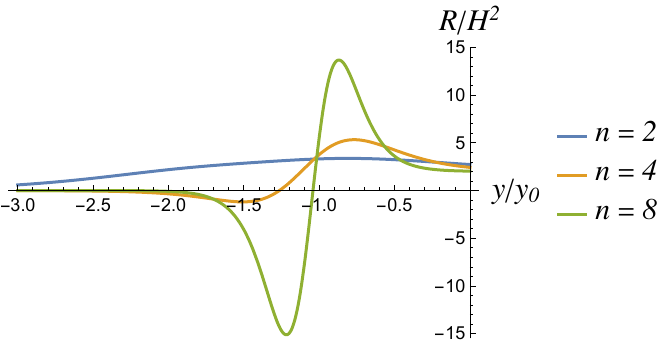}
\caption
{
 The behavior of the scalar curvature 
 $R/H^2$ as a function of $y/y_{0}$ for $n=2, 4, 8$.
 We adopted $c_{-}=1$, and $Ht=1$.
 }
  \label{fig:curvedirac}
\end{figure}
We now investigate how this profile impacts the bulk scalar curvature $R$. 
Fig.~\ref{fig:curvedirac} presents $R/H^2$
$y/y_{0}$
for various values of $n$.
As expected, the curvature becomes localized near 
$y=y_{0}$
and exhibits oscillatory features, including both positive and negative values. This further confirms that even in the bulk, the presence of a time-dependent edge modifies the geometry non-trivially.
\begin{figure}[H]
  \centering
  \begin{minipage}[b]{0.49\linewidth}
  \includegraphics[width=1\linewidth]{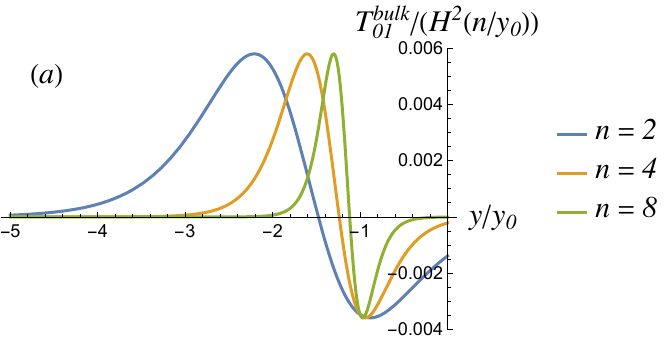}
  \end{minipage}\hspace{0.2cm}
  \begin{minipage}[b]{0.49\linewidth}
  \includegraphics[width=1\linewidth]{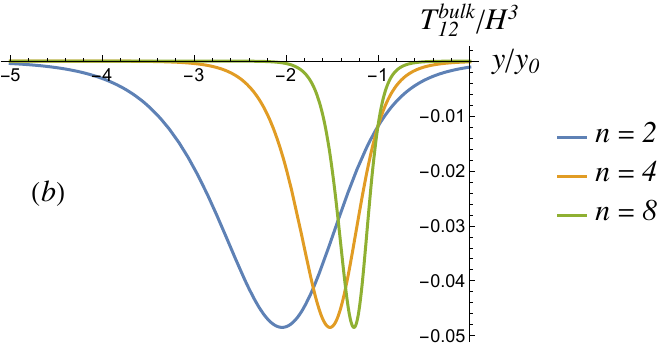}
  \end{minipage}
  \caption
{
 The behavior of the energy-momentum tensor in the bulk as a function of $y/y_{0}$ for $n=2, 4, 8$.
 Left and right panels correspond to 
 $T^{\text{bulk}}_{01}/(H^2(n/y_{0}))$ and 
 $T^{\text{bulk}}_{12}/H^3$.
 We adopted $c_{-}=1$, and $Ht=1$.
 }
  \label{fig:dirac0112}
\end{figure}
Next, we evaluate the components of the bulk energy-momentum tensor.
The panels in Fig.~\ref{fig:dirac0112} show the behavior of
$T^{\text{bulk}}_{01}$
and
$T^{\text{bulk}}_{12}$
for $n=2, 4, 8$, respectively.
The results are qualitatively similar to the Gaussian case,
where
$T^{\text{bulk}}_{01}$ 
includes regions of both positive and negative values, indicating the presence of direction-dependent energy transport parallel to the edge.
$T^{\text{bulk}}_{12}$
remains negative, representing a consistent momentum flow perpendicular to the edge.
As $n$ increases, the nonzero regions of these components become more sharply localized around the boundary $y=y_{0}$, indicating that the currents are effectively constrained to a narrow interface region. 
This is consistent with the anomaly-inflow mechanism, where edge anomalies are absorbed into the bulk.
In summary, the Fermi-Dirac profile yields similar qualitative behavior to the Gaussian case but differs quantitatively near the boundary. 
These results reinforce the robustness of the energy-momentum transfer between the edge and bulk regions under different smooth interpolation schemes.

\subsection{Quantized total energy flux in the bulk}
Finally, we discuss the physical meaning of the components
$T^{\text{bulk}}_{01}$
in terms of the conservation of energy-momentum in the entire system.
To do this, we focus on the behavior of the quantity 
$\int_{-\infty}^{0}dy T^{\text{bulk}}_{01}$
for both the Gaussian function and the Fermi-Dirac distribution function cases.
\begin{figure}[H]
  \centering
  \begin{minipage}[b]{0.49\linewidth}
  \includegraphics[width=1\linewidth]{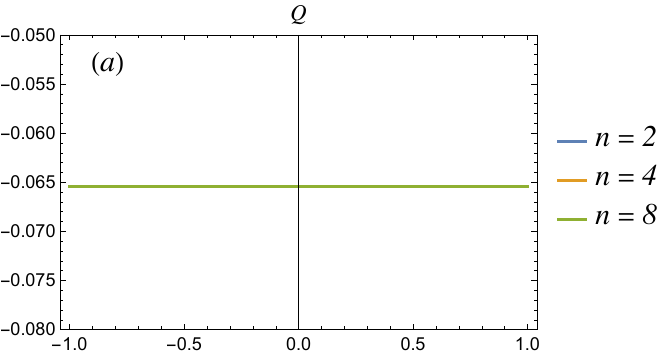}
  \end{minipage}\hspace{0.2cm}
  \begin{minipage}[b]{0.49\linewidth}
  \includegraphics[width=1\linewidth]{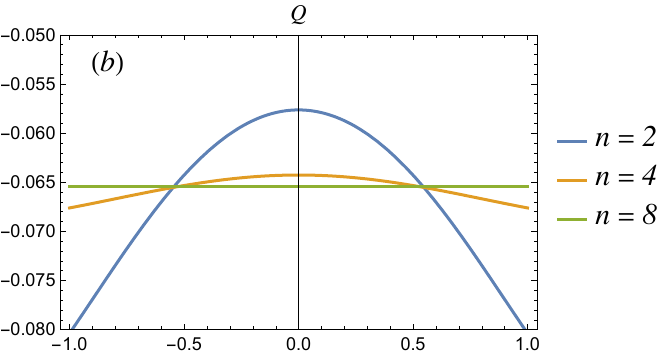}
  \end{minipage}
  \caption
{
 The behavior of the energy-momentum tensor in the bulk as a function of $-1<Ht<1$ for $n=2, 4, 8$.
 Left and right panels correspond to 
 $Q=H^{-2}\int_{-\infty}^{0} dy T^{\text{bulk}}_{01}$ for Gaussian function and Fermi-Dirac distribution function cases.
We adopted $c_{-}=1$.% and $x=0$.
}
  \label{fig:conserved}
\end{figure}
The panels in Fig.~\ref{fig:conserved} 
(a) and (b) present the time dependence of the quantity
$Q=\int_{-\infty}^{0}dy T^{\text{bulk}}_{01}$ for $n=2, 4, 8$, respectively.
%{\color{red}
This quantity $Q$ %naively 
represents the total energy flux in the bulk.%}
 %\textcolor{blue}{write sentences to declare that the total flux is the same with the uniform expansion with additional edge.}.
Fig.~\ref{fig:conserved} (a), for the Gaussian function case, 
the quantities are constant with respect to time, which means the conserved quantity.
Analytically, the quantity is computed as
\begin{align}
\int_{-\infty}^{0} dy
T^{\text{bulk}}_{01}
=
\left[
\frac{c_{-} H^2 \sec^4[H t] 
\left[
1-\cos [2 H t]
+e^{\left({y}/{{y_{0}}}\right)^n}(\cos [2 H t]-2)
\right]}
{48 \pi  
\left[
\tan ^2[H t]+e^{\left({y}/{y_{0}}\right)^n}
\right]^2}
\right]_{-\infty}^{0}
=
-\frac{c_{-}H^2}{48\pi}.
\end{align}
This quantity certainly does not depend on both the time and the quantity $n$.
%Furthermore, the result is consistent with the energy flux at late time derived in Sec.~III with different signs. 
Furthermore, the magnitude of this integrated bulk stress tensor $T_{01}^{\text{bulk}}$ is exactly the same energy flux at late time in \eqref{ppcomponent} derived in Sec.~III, but the sign is the opposite. % with different signs.
On the other hand, for the Fermi-Dirac distribution function case, 
%it seems that the quantity $Q$ depends on the quantity $n$. 
%In fact,
we can compute the integrated stress tensor analytically as
\begin{align}
\int_{-\infty}^{0} dy
T^{\text{bulk}}_{01}
=
\left[
-\frac{c_{-} H^2 \sec^4[H t] 
\left[
e^{n({y}/{y_{0}}-1)}(2-\cos [2 H t]) +1
\right]}
{48 \pi  
\left[
1+\tan^2[H t]+e^{n({y}/{y_{0}}-1)}
\right]^2}
\right]_{-\infty}^{0}
=
-\frac{c_{-} H^2}{48\pi}
\frac{\left[
1+e^{-n}(2-\cos [2 H t])\right]}
{ 
\left[
1+e^{-n}\cos^2[H t]
\right]^2}.
\end{align}
Thus, for $n\to \infty$ where $f'(0)$ vanishes, it is exactly the same as the Gaussian function case. %whereas for finite $n$, it depends on the time, which is not a conserved quantity.
%\textcolor{blue}{In this case, the boundary has nonvanishing extrinsic curvature and we expect there is also a contribution from the boundary terms given in terms of the extrinsic curvature.
%It is natural to expect that after including the boundary term the integrated momentum tensor $\int _{-\infty}^0T_{01}^{\text{bulk}}dy$ takes the quantized value but we postpone it to a future problem.}

%The origin of the $n$ dependence stems from the behavior of the boundary $y=0$.
%When the derivative 
%$f'(y)|_{y=0}$
%vanishes, the quantity $Q$
%integral $\int_{-\infty}^{0} dy T^{\text{bulk}}_{01}$ 
%is quantized.
%\textcolor{red}{
%However, when the derivative
%$f'(y)|_{y=0}$
%does not vanish, the quantity $Q$
%integral $\int_{-\infty}^{0} dy T^{\text{bulk}}_{01}$ 
%depends on the parameter $n$, i.e., it is not quantized.}
%\textcolor{green}{
%This is interpreted as being because, in this case, the bulk and boundary contributions are not clearly separated.
%This is because in this case the bulk and boundary contributions are not clearly separated.
%}
These results indicate that the sum of energy and momentum should be conserved throughout the system, even though the bulk is infinitely deep.
Therefore, the anomalous current on the edge cancels out with ``edge states'' ($T^{\text{bulk}}_{01}$ component) in the bulk, and there is no anomaly in the entire system of bulk and edge.
Thus, the quantity $T^{\text{bulk}}_{01}$ is interpreted as the energy flux (or equivalently momentum flux $T^{\text{bulk}}_{10}$ because of the symmetric tensor) density in the bulk. 
%{\color{red}This suggests that the Nielsen-Ninomiya theorem ``like''.}
The integral of $T_{01}^{\text{bulk}}$ is quantized, but the distribution of $T_{01}^{\text{bulk}}$ itself depends on the details of the metric.
Indeed, $T_{01}^{\text{bulk}}$ can take both positive and negative values.

%{\color{red}
%What is the relation of 
%\begin{itemize}
%    \item Hall viscosity?
%    \item Bulk viscosity?
%\end{itemize}
%}

\section{Conclusion}
As one of the most well-understood and experimentally accessible topological phases, quantum Hall systems have the potential to serve as a natural testing ground for exploring both fundamental and practical aspects. 
In this study, we investigated the energy fluxes at both the edge and in the bulk of a quantum Hall system with the expanding edge, within the framework of gravitational anomalies.
%We first clarified that the appropriate formulation of the gravitational anomaly in this context is the covariant anomaly, which transforms covariantly under coordinate transformations.
Given that quantum Hall systems are fundamentally defined by their edge states arising from bulk topology, we analyzed the gravitational anomaly-inflow mechanism %not only at the edge but also in the bulk.
including the bulk gravitational Chern-Simons term.
%gravitational anomalies at the edge.
%not only at the edge but also in the bulk.
As a key result, we analytically demonstrated that the edge gravitational anomaly corresponds to a covariant anomaly, arising from the anomaly-inflow of the conserved bulk current to the edge.
Using this covariant anomaly framework, we computed the energy fluxes in two ways—solving for the covariant gravitational anomaly directly and focusing on the Weyl anomaly.
We showed that they emerge along the edge in the late time limit, analogous to Hawking radiation.
Furthermore, our analysis of the bulk gravitational anomaly revealed the existence of a non-trivial energy flux component in the bulk region due to the conservation of energy and momentum in the entire system.

In the present paper, we focused on the behavior of the energy-momentum tensor in $(2+1)$~dimensions, including the bulk as well as the expanding edge region, as in previous studies~\cite{Hotta2022, Nambu2023, Yoshimoto2025}.
We found that the total energy flux parallel to the edge is quantized.
Its magnitude equals that of the edge flux, but the sign is opposite.
On the other hand, the role of the bulk shear stress tensor $T^{\text{bulk}}_{12}$ is unclear.
Revealing the physical meaning of 
$T^{\text{bulk}}_{12}$ is needed for further study.
It is also intriguing to understand the properties of entanglement entropy~\cite{Hughes2016} in this quantum Hall system and discuss the information loss problem.
It may also be interesting to consider the case of the non-equilibrium property of the quantum Hall system with an expanding edge, such as the time-dependent drift velocity case.
Moreover, it is important to clarify how to measure the temperature of this quantum Hall system experimentally, e.g., to identify the Hawking temperature.
These issues are left for future studies.

\acknowledgments
We thank Yasusada Nambu, Takashi Oka, Kohei Kawabata, Daichi Nakamura, Akihiro Ozawa, Go Yusa, Kazuya Yonekura, Yunhyeon Jeong, and Masahiro Hotta for useful discussions.
Y.S. was supported by JSPS KAKENHI Grant-in-Aid for Research Activity Start-up (Grant No. 24K22854) and Grant No. 25KJ0065.
 TN is supported by MEXT KAKENHI Grant No.~	23K13094, 24H00944 and JST PRESTO Grant No.~JPMJPR2359.

\begin{appendix}
\section{Formulas in Minkowski and null coordinates cases}
In this Appendix, we summarize the formulas 
(covariant totally anti-symmetric tensor $\bar{\epsilon}_{ij}$,
Christoffel symbols
$\Gamma^{i}_{jk}$, and
scalar curvature $R$)
in $(1+1)$-dimensional conformally flat spacetime, where the flat spacetime is the case for both Minkowski coordinates
$(x^{0}, x^{1})$ 
and null coordinates $(x^{+}, x^{-})$.
\subsection{Minkowski coordinate case}
The line element is 
\begin{align}
ds^2=\omega^2(-(dx^{0})^2+(dx^{1})^2)
=h_{ij}dx^{i}dx^{j},
\label{minkowski}
\end{align}
where we introduced the metric 
$h_{ij}$ in $(1+1)$-dimension as
\begin{align}
h_{ij}=
\omega^2
\begin{pmatrix}
-1&&0\\
0&&1
\end{pmatrix},
\quad
h^{ij}=
\omega^{-2}
\begin{pmatrix}
-1&&0\\
0&&1
\end{pmatrix}.
\end{align}
The covariant totally anti-symmetric tensor is defined by
\begin{align}
\bar{\epsilon}_{ij}=\sqrt{-h}\epsilon_{ij}, 
\quad
\bar{\epsilon}^{ij}=\frac{\epsilon^{ij}}{\sqrt{-h}}
\end{align}
with $h:=\det h_{ij}=-\omega^4$.
The components are explicitly given by
\begin{align}
\bar{\epsilon}_{01}=-\bar{\epsilon}_{10}=-\omega^2,
\quad
\bar{\epsilon}^{01}=-\bar{\epsilon}^{10}=\omega^{-2}.
\end{align}
In this coordinate, the non-zero components of the Christoffel symbols and scalar curvature are
\begin{align}
\Gamma^{0}_{00}
=
\Gamma^{0}_{11}
=
\Gamma^{1}_{01}
=
\Gamma^{1}_{10}
=
\frac{\partial_{0}{\omega}}{\omega},
\quad
\Gamma^{0}_{01}
=
\Gamma^{0}_{10}
=
\Gamma^{1}_{00}
=
\Gamma^{1}_{11}
=
\frac{\partial_{1}{\omega}}{\omega},
\quad
R=
\frac{2}{\omega^2}(\partial^2_{0}-\partial^2_{1})\log\omega
\end{align}

\subsection{null coordinate case} 
Introducing the null coordinate,
$x^{\pm}:=x^{0}\pm x^{1}$,
the line-element~\eqref{minkowski} can be rewritten as
\begin{align}
ds^2=-\omega^2dx^{+}dx^{-}
=h_{ij}dx^{i}dx^{j},
\end{align}
where, in this coordinate, the metric tensor is defined as
\begin{align}
h_{ij}=
-\frac{\omega^2}{2}
\begin{pmatrix}
0&&1\\
1&&0
\end{pmatrix},
\quad
h^{ij}=
-2\omega^{-2}
\begin{pmatrix}
0&&1\\
1&&0
\end{pmatrix}.
\end{align}
The covariant totally anti-symmetric tensor
$\bar{\epsilon}_{ij}=\sqrt{-h}\epsilon_{ij}$
and
$\bar{\epsilon}^{ij}={\epsilon^{ij}}/{\sqrt{-h}}$
with $\sqrt{-h}=\omega^2/2$
are related to the general coordinate transformation
\begin{align}
\bar{\epsilon}_{+-}=-\bar{\epsilon}_{-+}=\frac{\omega^2}{2},
\quad
\bar{\epsilon}^{+-}=-\bar{\epsilon}^{-+}=-\frac{2}{\omega^{2}}.
\end{align}
The non-zero components of the Christoffel symbols and scalar curvature are
\begin{align}
\Gamma^{+}_{++}
=
\frac{2\partial_{+}{\omega}}{\omega},
\quad
\Gamma^{-}_{--}
=
\frac{2\partial_{-}{\omega}}{\omega},
\quad
R
=\frac{8}{\omega^2}\partial_{+}\partial_{-}\log\omega.
\end{align}

\section{Detailed computation of QHS with an expanding edge}
We find the conditions for a continuous and smooth connection with region $\rm{II}$ at 
$x=x_{\rm{I}}=L/2$ and $x=x_{\rm{III}}=-L/2$, respectively.
Firstly, at $x=x_{\rm{I}}=L/2$, we obtain the following condition
\begin{align}
x^{+}_{\rm{I}}[t+L/2]-x^{-}_{\rm{I}}[t-L/2]=L,
\label{cont}
\end{align}
where its derivative with respect to the time $t$ reads
\begin{align}
\frac{dx^{+}_{\rm{I}}}{dx^{+}}[t+L/2]
-\frac{dx^{-}_{\rm{I}}}{dx^{-}}[t-L/2]=0
\label{smooth}
\end{align}
with $d/dt=d/dx^{+}=d/dx^{-}$.
Combining Eq.~\eqref{smooth} and the condition where the metric is connected at $x=x_{\rm{I}}=L/2$, the conformal factor 
$e^{\Theta(t)}$
satisfies
%\begin{align}
%\frac{dx^{+}_{\rm{I}}}{dx^{+}}[t+L/2]
%=\frac{dx^{-}_{\rm{I}}}{dx^{-}}[t-L/2]
%=e^{\Theta(t)}.
%\end{align}
%Shifting the argument of the conformal factor as $t\to t+x-L/2=x^{+}-L/2$, 
%and
%$t\to t-x+L/2=x^{-}+L/2$
\begin{align}
\frac{dx^{+}_{\rm{I}}}{dx^{+}}[x^{+}]
=
e^{\Theta(x^{+}-L/2)},
%\quad
%\frac{dx^{-}_{\rm{I}}}{dx^{-}}[x^{-}]
%=
%e^{\Theta(x^{-}+L/2)}
\end{align}
where we shifted the argument of the conformal factor as $t\to t+x-L/2=x^{+}-L/2$.
Thus the null coordinate in region I, 
$x^{+}_{\rm{I}}$,  
becomes a function of $x^{+}$ as
\begin{align}
x^{+}_{{\rm{I}}}[x^{+}]
=
\int_{0}^{x^{+}}dye^{\Theta(y-L/2)}
=
\int_{-L/2}^{x^{+}-L/2}dy^{\prime}e^{\Theta(y^{\prime})}
=
\Phi[x^{+}-L/2]-\Phi[-L/2],
\end{align}
where, in the second equality, we changed the integral variable as $y\to y^{\prime}=y-L/2$, and, in the third equality, we introduced the function $\Phi[x]$ explicitly given by Eq.~\eqref{phi}.
The coordinate $x^{-}_{\rm{I}}[x^{-}]$ is determined by Eq.~\eqref{cont}
\begin{align}
x^{-}_{\rm{I}}[x^{-}]
=x^{+}_{\rm{I}}[x^{-}+L]-L
=\Phi[x^{-}+L/2]-\Phi[-L/2]-L,
\end{align}
where in the first equality, we shifted the argument as
$t\to t-x +L/2=x^{-}+L/2$.
Thus, in region II, the metric between region I and II relates as
\begin{align}
ds^2_{\rm{II}}
=-e^{2\Theta(t)}dx^{+}dx^{-}
=-\exp
\left[
2\Theta(t)-\Theta(x^{+}-L/2)-\Theta(x^{-}+L/2)
\right]dx^{+}_{\rm{I}}dx^{-}_{\rm{I}}.
\end{align}
At the boundary, $x^{+}=t+L/2$ and $x^{-}=t-L/2$, the conformal factor becomes unity.

The similar procedure above yields the following conditions at $x=x_{\rm{III}}=-L/2$
%%%%%%%%%%%%%%%%%%%%%%%%%%%%%%%%%%%%%%%%%%%%
\if0
\begin{align}
x^{+}_{\rm{III}}[t-L/2]-x^{-}_{\rm{III}}[t+L/2]=-L,
\end{align}
\begin{align}
\frac{dx^{+}_{\rm{III}}}{dx^{+}}[t-L/2]
-\frac{dx^{-}_{\rm{III}}}{dx^{-}}[t+L/2]=0,
\end{align}
\begin{align}
\frac{dx^{+}_{\rm{III}}}{dx^{+}}[t-L/2]
=\frac{dx^{-}_{\rm{III}}}{dx^{-}}[t+L/2]
=e^{\Theta(t)}
\end{align}
Shifting the argument as $t\to t+x+L/2=x^{+}+L/2$
and
$t\to t-x-L/2=x^{-}-L/2$
\begin{align}
\frac{dx^{+}_{\rm{III}}}{dx^{+}}[x^{+}]
=
e^{\Theta(x^{+}+L/2)},
\quad
\frac{dx^{-}_{\rm{III}}}{dx^{-}}[x^{-}]
=
e^{\Theta(x^{-}-L/2)}
\end{align}
\fi
%%%%%%%%%%%%%%%%%%%%%%%%%%%%%%%%%%%%%%%%%%%%
\begin{align}
x^{+}_{\rm{III}}[x^{+}]
%=
%\int_{0}^{x^{+}}dy e^{\Theta(y+L/2)}
%=
%\int_{L/2}^{x^{+}+L/2}dy e^{\Theta(y)}
=\Phi[x^{+}+L/2]-\Phi[L/2]
\end{align}
for
$x^{+}_{\rm{III}}$
and
\begin{align}
x^{-}_{\rm{III}}[x^{-}]
=
x^{+}_{\rm{III}}[x^{-}-L]+L
=\Phi[x^{-}-L/2]-\Phi[L/2]+L
\end{align}
for $x^{-}_{\rm{III}}$.
%where in the first equality, we shifted the argument as $t\to t-x -L/2=x^{-}-L/2$.
Thus, the metric of the two regions II and III relates to 
\begin{align}
ds^2_{\rm{II}}
=-e^{2\Theta(t)}dx^{+}dx^{-}
=-\exp
\left[
2\Theta(t)-\Theta(x^{+}+L/2)-\Theta(x^{-}-L/2)
\right]dx^{+}_{\rm{III}}dx^{-}_{\rm{III}}
\end{align}
At the boundary, $x^{+}=t-L/2$ and $x^{-}=t+L/2$, the conformal factor also becomes unity.

\section{Proof of the equivalence of Eqs.~\eqref{ppcomponent} and~\eqref{weylppcomponentc}}
Here, we analytically demonstrate the equivalence between Eqs.~\eqref{ppcomponent} and~\eqref{weylppcomponentc}.
To do so, we analyze the Schwarzian derivative $\text{Sch}[f, x^{+}_{\rm{III}}]$ appearing in Eq.~\eqref{weylppcomponentc}.
Let us begin by considering the coordinate transformation function 
$f[x^{+}_{\rm{III}}]$, given by
\begin{align}
f[x^{+}_{\rm{III}}]
=
\Phi[-L+\Phi^{-1}[x^{+}_{\rm{III}}+\Phi[L/2]]]-\Phi[-L/2]
=\Phi[g[x^{+}_{\rm{III}}]]-\Phi[-L/2]
\end{align}
where we introduced a function $g[x^{+}_{\rm{III}}]$
\begin{align}
g[x^{+}_{\rm{III}}]
=-L+\Phi^{-1}[x^{+}_{\rm{III}}+\Phi[L/2]].
\end{align}
Since 
$f[x^{+}_{\rm{III}}]$
is a composition of functions, we can compute the Schwarzian derivative 
$\text{Sch}[f, x^{+}_{\rm{III}}]$ 
using the chain rule:
\begin{align}
\text{Sch}[f, x^{+}_{\rm{III}}]
&=
\frac{f^{\prime\prime\prime}}{f^{\prime}}
-\frac{3}{2}\left(\frac{f^{\prime\prime}}{f^{\prime}}\right)^2
=
\left[
\frac{\partial^3\Phi/\partial g^3}{\partial\Phi/\partial g}
-\frac{3}{2}
\left(
\frac{\partial^2\Phi/\partial g^2}{\partial\Phi/\partial g}
\right)^2
\right]
\left(
g^{\prime}
\right)^2
+\frac{g^{\prime\prime\prime}}{g^{\prime}}
-\frac{3}{2}\left(\frac{g^{\prime\prime}}{g^{\prime}}\right)^2,
\label{composition}
\end{align}
where the derivatives denoted by
$\ ^{\prime}$ 
are taken with respect to
$x^{+}_{\rm{III}}$.
Using the function $\Phi[x]$ defined by Eq.~\eqref{eq:Phi}, the first term of Eq.~\eqref{composition} in the second equality is represented by 
$g[x^{+}_{\rm{III}}]$
as follows
\begin{align}
\frac{\partial^3\Phi/\partial g^3}{\partial\Phi/\partial g}
-\frac{3}{2}
\left(
\frac{\partial^2\Phi/\partial g^2}{\partial\Phi/\partial g}
\right)^2
=
\frac{H^2(1+\cos^2(g))}{2\cos^2(g)}.
\end{align}
For the second term of Eq.~\eqref{composition}, we evaluate the combination of derivatives of $g[x^{+}_{\rm{III}}]$.
From the structure of $g[x^{+}_{\rm{III}}]$, we use the relations:
$(g^{\prime\prime}/g^{\prime})^2
=1-\left(g^{\prime}\right)^2$
and
$g^{\prime\prime\prime}/g^{\prime}
=1-2\left(g^{\prime}\right)^2$,
which yields
\begin{align}
\frac{g^{\prime\prime\prime}}{g^{\prime}}
-\frac{3}{2}\left(\frac{g^{\prime\prime}}{g^{\prime}}\right)^2
=-\frac{H^2\left(1+\left(g^{\prime}\right)^2\right)}{2}.
\end{align}
Combining both contributions, the Schwarzian derivative becomes:
\begin{align}
\text{Sch}[f, x^{+}_{\rm{III}}]
=
\frac{\left(g^{\prime}\right)^2}{2}
\frac{H^2(1+\cos^2(g))}{\cos^2(g)}
-\frac{H^2\left(1+\left(g^{\prime}\right)^2\right)}{2}
=
\frac{H^2}{2}
\left[
\left(\frac{g^{\prime}}{\cos(g)}\right)^2-1\right].
\end{align}
On the other hand, Eq.~\eqref{ppcomponent} involves the quantity
$1-\left(f^{\prime}\right)^2$,
which can be expressed using the chain rule as:
\begin{align}
1-\left(f^{\prime}\right)^2
=
1-\left(\frac{\partial \Phi}{\partial g}\right)^2\left(g^{\prime}\right)^2
=
1-
\left(
\frac{g^{\prime}}{\cos(g)}
\right)^2.
\end{align}
Hence, we can relate the Schwarzian derivative to Eq.~\eqref{ppcomponent} via:
\begin{align}
-\frac{2\text{Sch}[f, x^{+}_{\rm{III}}]}{H^2}
=1-\left(
\frac{\partial f[x^{+}_{\rm{III}}]}
{\partial x^{+}_{\text{III}}}
\right)^2.
\end{align}
This clearly shows that Eqs.~\eqref{ppcomponent} and~\eqref{weylppcomponentc} are indeed equivalent, as both describe the same geometric transformation encoded via the function $f[x^{+}_{\rm{III}}]$.

\section{Explicit form of the bulk energy-momentum tensor}
Here, we present the explicit form of the bulk energy-momentum tensor 
$T^{\text{bulk}}_{01}$ and  $T^{\text{bulk}}_{12}$.
The spatial profile function $f(y)$ is chosen in Eq.~\eqref{gauss} for the Gaussian function type case and Eq.~\eqref{Dirac} for the Fermi-Dirac distribution function type case.
\begin{itemize}
 \item{Gaussian function case}
    \begin{align}
    T^{\text{bulk}}_{01}
    &=
    \frac{c_{-} H^2 (n/y_{0}) \left({y}/{{y_{0}}}\right)^n 
    \sec^4[H t] e^{\left({y}/{y_{0}}\right)^n} 
    \left[2 \sin ^2[H t] 
    \left(\tan ^2[H t]+e^{\left({y}/{y_{0}}\right)^n}\right)
    +
    \left(e^{\left({y}/{y_{0}}\right)^n}
    -5 \tan^2[H t]\right)
    \right]}
    {48\pi (y/y_{0}) 
    \left[
    \tan ^2[H t]+e^{\left({y}/{y_{0}}\right)^n}
    \right]^3},
   \\
   \quad
    T^{\text{bulk}}_{12}
    &=
    \frac{c_{-} H^3 \tan [H t] \sec^4[H t] 
    e^{\left({y}/{y_{0}}\right)^n} 
    \left[e^{\left({y}/{y_{0}}\right)^n}-1\right]
    \left[
    \cos [2 H t]
    \left(e^{\left({y}/{y_{0}}\right)^n}-1\right)-5 e^{\left({y}/{y_{0}}\right)^n}+1
    \right]}
    {24\pi
    \left[\tan ^2[H t]
    +e^{\left({y}/{y_{0}}\right)^n}
    \right]^4},
    \end{align}
 \item Fermi-Dirac distribution function case
    \begin{align}
    T^{\text{bulk}}_{01}
    &=
    \frac{c_{-} H^2 (n/y_{0}) 
    \sec^6[H t] e^{n \left({y}/{y_{0}}-1\right)}
    \left[
    2 \cos [2 H t] 
    \left(e^{n \left({y}/{y_{0}}-1\right)}+4\right)
    -\cos [4 H t] e^{n
    \left({y}/{y_{0}}-1\right)}+3 e^{n \left({y}/{y_{0}}-1\right)}-4
    \right]}
    {192\pi
    \left[\tan^2[H t]+e^{n \left({y}/{y_{0}}-1\right)}+1\right]^3},
   \\
   \quad
    T^{\text{bulk}}_{12}
    &=
    \frac{c_{-} H^3 \tan [H t] 
    \sec^4[H t] e^{n \left({y}/{y_{0}}-1\right)} 
    \left[
    e^{n \left({y}/{y_{0}}-1\right)}+1
    \right]
    \left[
    \cos [2 H t] e^{n
    \left({y}/{y_{0}}-1\right)}
    -5 e^{n \left({y}/{y_{0}}-1\right)}-4
    \right]}
    {24\pi
    \left[
    \tan^2[H t]+e^{n 
    \left({y}/{y_{0}}-1\right)}+1
    \right]^4}.
    \end{align}
\end{itemize}

\end{appendix}

\bibliography{ref}

\end{document}